\documentclass[%
 reprint,
superscriptaddress,
 amsmath,amssymb,
 aps,
floatfix,
]{revtex4-2}
\usepackage[caption=false]{subfig}
\usepackage{graphicx}
\usepackage{dcolumn}
\usepackage{bm}
\begin{document}
\preprint{APS/123-QED}

\title{Characterization and suppression of background light shifts in an optical lattice clock}

\author{R.J. Fasano}
\author{Y.J. Chen}
\author{W.F. McGrew}
\author{W.J. Brand}
\affiliation{National Institute of Standards and Technology, 325 Broadway, Boulder, Colorado 80305, USA
}
\affiliation{University of Colorado, Department of Physics, Boulder, Colorado 80309, USA}

\author{R.W. Fox}
\affiliation{National Institute of Standards and Technology, 325 Broadway, Boulder, Colorado 80305, USA
}

\author{A.D. Ludlow}
\email{andrew.ludlow@nist.gov}
\affiliation{National Institute of Standards and Technology, 325 Broadway, Boulder, Colorado 80305, USA
}
\affiliation{University of Colorado, Department of Physics, Boulder, Colorado 80309, USA}

\date{\today}

\begin{abstract}
Experiments involving optical traps often require careful control of the ac Stark shifts induced by strong confining light fields. By carefully balancing light shifts between two atomic states of interest, optical traps at the magic wavelength have been especially effective at suppressing deleterious effects stemming from such shifts.  Highlighting the power of this technique, optical clocks today exploit Lamb-Dicke confinement in magic-wavelength optical traps, in some cases realizing shift cancellation at the ten parts per billion level. Theory and empirical measurements can be used at varying levels of precision to determine the magic wavelength where shift cancellation occurs.  However, lasers exhibit background spectra from amplified spontaneous emission or other lasing modes which can easily contaminate measurement of the magic wavelength and its reproducibility in other experiments or conditions. Indeed, residual light shifts from laser background have plagued optical lattice clock measurements for years. In this work, we develop a simple theoretical model allowing prediction of light shifts from measured background spectra. We demonstrate good agreement between this model and measurements of the background light shift from an amplified diode laser in an Yb optical lattice clock. Additionally, we model and experimentally characterize the filtering effect of a volume Bragg grating bandpass filter, demonstrating that application of the filter can reduce background light shifts from amplified spontaneous emission well below the $10^{-18}$ fractional clock frequency level. This demonstration is corroborated by direct clock comparisons between a filtered amplified diode laser and a filtered titanium:sapphire laser.
\end{abstract}
\maketitle

\section{\label{sec:level1}Introduction}
Optical traps allow for strong confinement of atomic systems, finding a multitude of applications in atomic clocks \cite{katori2003ultrastable}, quantum simulation \cite{bloch2008many}, and quantum information \cite{briegel2000quantum}. However, the ac Stark interaction induced by the trapping laser generally leads to undesired shifts, broadening, and decoherence mechanisms that otherwise compromise precise measurement or control of the atomic system. Many of these effects can be reduced with an optical trap operating at a magic wavelength, where two electronic states of interest are shifted equally and the transition energy between them remains unperturbed. This technique has achieved dramatic success for precision spectroscopy in optical clocks based on optical lattices \cite{takamoto2005optical, ye2008quantum, barber2008optical, kohno2009one, westergaard2011lattice} or optical tweezers \cite{saskin2019narrow}. Semi-empirical and ab initio models can reasonably predict the existence and value of a magic wavelength for a given atomic system \cite{ovsyannikov2006polarisation, mitroy2010theory}. Later, more precise determination of the magic wavelength is afforded by empirical measurement and enhanced theoretical models, with care required in accounting for higher-order shifts from the light field \cite{brusch2006hyperpolarizability, katori2009magic, bloom2014optical, brown2017hyperpolarizability, porsev2018multipolar, ushijima2018operational, nemitz2019modeling}.  However, experimental determinations of the magic wavelength are strictly applicable only to the laser utilized in a particular experimental apparatus.  This is because real laser systems exhibit background spectral content away from the magic wavelength due to effects such as amplified spontaneous emission (ASE) or multi-mode lasing.  Such a spectrum can introduce a large and potentially fluctuating Stark shift that is generally difficult to characterize or control, yielding a significant shift in the wavelength at which zero differential light shift is observed compared to an ideal, background-free laser system.  Moreover, this shift would vary with experimental conditions, such as laser and optical systems employed. Additionally, such shifts may not be detected in internal clock comparisons that share an optical lattice laser, leading to surprising results when clocks in different laboratories are compared.

This caution is especially relevant to diode laser systems that employ tapered amplifiers (TA), whose large, broadband ASE spectra have caused experimental challenges for clock operation \cite{le2012comparison, le2013experimental, koller2017transportable, bloom2014building}. Indeed, background-induced light shifts at the Hz level ($10^{-15}$ fraction of the clock frequency) or much higher are easily possible. There have been various attempts to mitigate ASE effects through amplifier optimization \cite{zhou2018characterization, voigt2001characterization}, spatial filtering with optical fibers \cite{bolpasi2010double, nyman2006tapered}, or spectral filtering using volume Bragg gratings \cite{brown2017hyperpolarizability, koller2017transportable}, Fabry-Perot etalons \cite{bienaime2012fast}, or absorption cells \cite{dumke2002micro}. For an additional review of the origins and consequences of ASE, as well as an in-depth investigation of several strategies for ASE characterization, see \cite{zhou2018characterization}. Despite these efforts, the risk of lingering Stark shifts is often sufficiently great that current best practice is to employ filtered titanium:sapphire (Ti:S) lasers for the optical trap, benefiting from the significantly weaker spectral background. Nevertheless, the measurement of optical spectral purity of any laser can be challenging at the levels required for optical clocks today and in the future. Here we present a detailed measurement of a laser spectrum across eight decades of frequency, offering a rigorous evaluation of residual light shifts stemming from the laser background and supporting optical clock performance well beyond the current state-of-the-art. Furthermore, we show that with appropriate precautions, amplified diode systems can be employed in optical clocks while still limiting residual light shifts to below the $10^{-18}$ fractional frequency level. This is especially relevant for the many scientific and technological applications requiring fieldable or portable optical clocks and experiments \cite{koller2017transportable, takamoto2020test}, including space-based optical clocks \cite{origlia2018towards}, where the reduced size, weight, power consumption, and complexity of diode lasers relative to Ti:S systems is appealing. 

In Section \ref{sec:shifts} of this work, we provide a simple theoretical model to predict light shifts from observed background spectra. We apply this model to predict large shifts arising from an amplified diode laser, finding good agreement with direct atomic measurements in an Yb optical lattice clock. In Section \ref{sec:filtering}, we provide a brief overview and experimental measurements of spectral filtering using volume Bragg gratings. The characterized ASE spectra are used in conjunction with the measured grating transfer function to predict a residual background light shift of $8\times 10^{-21}$ under typical experimental conditions. In Section \ref{sec:uncertainty}, we assess the uncertainty in this estimate arising from several effects such as optical characterization limits and fluctuating shifts, conservatively constraining its magnitude to $5\times 10^{-20}$. Finally, in Section \ref{sec:experimental_validation}, we corroborate this uncertainty with a direct atomic measurement of the filtered background light shift, constraining it at or below the $1\times 10^{-18}$ level.

\section{\label{sec:shifts}Background light shifts}
We assume that the laser spectrum can be described by two components: a spectrally-narrow ``carrier'' containing power $P$ at a frequency $\nu_{trap}$ and a broadband background described by the power spectral density $S(\nu)$. Considering only effects from electric dipole polarizability, light shifts from the two components can be expressed
\begin{equation}\label{eq:trap_shift}
\Delta\nu_{trap} = -\frac{P}{\epsilon_0 hc\pi w_0^2}\Delta\alpha(\nu_{trap})
\end{equation}
and
\begin{equation}\label{eq:E1_shift}
    \Delta\nu_{background} = -\int_0^\infty\frac{1}{\epsilon_0 hc\pi w_0^2}\Delta\alpha(\nu)S(\nu)W(\nu)d\nu,
\end{equation}
where $\Delta\alpha(\nu)=\text{Re}[\alpha_{(e)}(\nu)-\alpha_{(g)}(\nu)]$ is the differential polarizability between the two clock states, $\epsilon_0$ is the permittivity of free space, $h$ is Planck's constant, $c$ is the speed of light in vacuum, and $w_0$ is the beam diameter at $1/e^2$ intensity. Typically, $\nu_{trap}$ is chosen to be the magic frequency $\nu_{magic}$ such that the shift given in (\ref{eq:trap_shift}) vanishes. The factor $W(\nu)$ in (\ref{eq:E1_shift}) arises due to dephasing of the ASE relative to the standing wave of the optical lattice and is discussed in Appendix \ref{sec:incoherence}. This weighting factor converges to 1 at $\nu=\nu_{magic}$ and oscillates around 1/2 elsewhere. As the divergence from 1/2 is only significant near the magic wavelength, where the differential polarizability is small, we will assume $W(\nu)=1/2$ for all calculations.
\\
\\
The E1 polarizabilities of the ground and excited states can be calculated using known atomic state parameters:
\begin{equation}\label{eq:harmonic_polarizability}
    \alpha(\nu)_{(j)}=\frac{3\epsilon_0 c^3}{(2\pi)^3}  \sum_i \frac{\Gamma_{ij}f_{ij}}{\nu_{ij}^4-\nu^2\nu_{ij}^2-2\pi i\Gamma_{ij}\nu^3},
\end{equation}
where the sum is taken over all intermediate states coupled to the state of interest and $j$ represents either the ground or excited state. Here, $\Gamma_{ij}$, $f_{ij}$, and $\nu_{ij}$ are the linewidth, branching ratio, and frequency of the transition from the ground state $|g\rangle$ or the excited state $|e\rangle$ to the intermediate state $|i\rangle$.

\begin{figure}[t!]
\includegraphics[width=8.6cm]{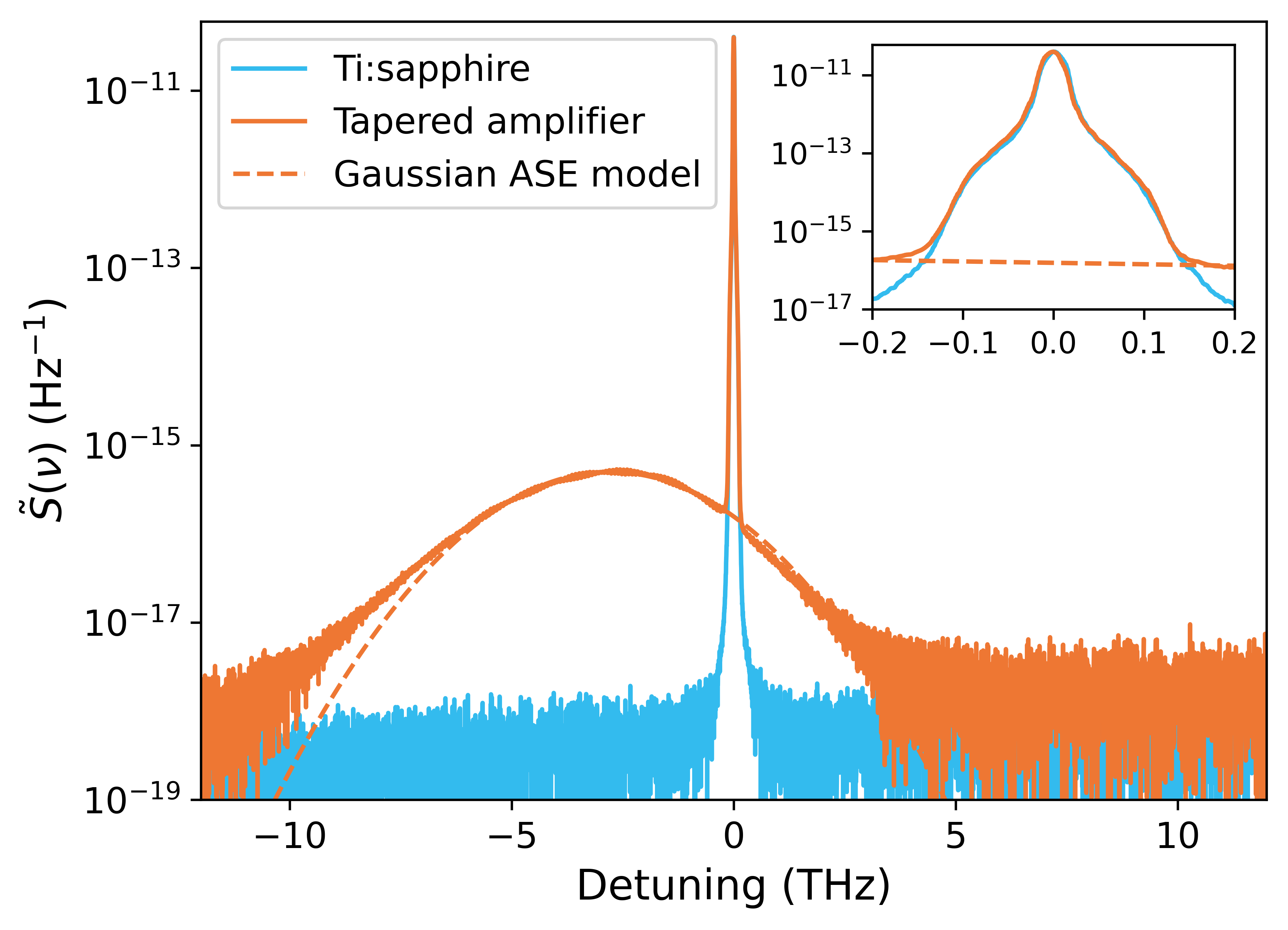}
\caption{\label{fig:osa-measurements} Optical spectrum analyzer measurements of background spectra of the Ti:S and TA systems as a function of detuning relative to the magic wavelength. Near the peak, the achievable resolution is limited by the optical rejection of the OSA, shown in the inset.}
\end{figure}

For the carrier at the magic wavelength, it is convenient to define a dimensionless trap depth given by a ratio relative to the photon recoil energy of the lattice laser, $E_r=h^2\nu_{trap}^2/2mc^2$. The dimensionless trap depth parameter can be experimentally determined with motional sideband spectroscopy:
\begin{equation}
    U = \frac{P}{\epsilon_0 c \pi w_0^2E_r}\alpha(\nu_{magic})=\bigg (\frac{h\nu_z}{2E_r}\bigg )^2,
\end{equation}
where $\nu_z$ is the longitudinal trap frequency. The fractional background shift per unit frequency, or shift spectral density, can be expressed as the product of dimensionless parameters separately describing the trap, the background spectrum, and the atomic response, yielding a total shift
\begin{equation}\label{eq:fractional_shift}
    \frac{\Delta\nu_{clock}}{\nu_{clock}} = -\frac{1}{2}U\int_0^\infty\tilde S(\nu)\Delta\alpha_{E1}'(\nu)d\nu,
\end{equation}
where 
\begin{equation}\label{eq:delta_alpha}
    \Delta\alpha'_{E1}(\nu)=\frac{E_r}{h\nu_{clock}}\frac{\Delta\alpha(\nu)}{\alpha(\nu_{magic})}
\end{equation}
and $\tilde S(\nu)=S(\nu)/P$ is the background power spectral density relative to the peak power. 
\\
\\
Using an optical spectrum analyzer (OSA), we measure the background spectra of a Ti:S and diode-seeded tapered amplifier (Fig. \ref{fig:osa-measurements}) \footnote{Equipment used for these measurements includes: Ti:S laser (M-Squared SolsTiS), amplified diode laser (Toptica TA Pro), and optical spectrum analyzer (Yokogawa AQ6373B). Product information is provided for informational purposes and does not represent an endorsement by NIST.}. The Ti:S exhibits spectral purity down to the noise floor of the measurement, while the TA possesses an evident broadband background spectrum which is well fit by a Gaussian model,
\begin{equation}\label{eq:gaussian-profile}
    \tilde S(\nu)=\tilde S_0\exp\bigg (-\frac{(\nu-\nu_{ASE})^2}{2\sigma_{ASE}^2} \bigg )
\end{equation}
The parameters of this profile, and thus the resulting background light shift, can be tuned with the amplifier current and temperature. Except where specified otherwise, all measurements in this work are taken at an operational setpoint of 3210 mA and 20 $^{\circ}$C, at which the spectrum is described by $\nu_{ASE}=\nu_{magic}-2.8\text{ THz}$, $\tilde S_0=4.7\times 10^{-16} \text{ Hz}^{-1}$, and $\sigma_{ASE}=2.1\text{ THz}$. 

Together with this empirically-observed Gaussian profile, we can obtain a closed-form estimate of the shift from (\ref{eq:fractional_shift}) by making a linear approximation of the differential polarizability as
\begin{equation}
    \Delta\alpha'_{E1}=\frac{d\Delta\alpha'_{E1}}{d\nu}(\nu-\nu_{magic})
\end{equation}
with $d\Delta\alpha'_{E1}/d\nu=4.5(3)\times 10^{-26}\text{ Hz}^{-1}$ \cite{brown2017hyperpolarizability}. This approximation agrees with (\ref{eq:delta_alpha}) to better than 5\% within the typical full-widths at half-max of ASE spectra observed in this work. Under these two approximations, the shift is
\begin{equation}\label{eq:shift_approx}
    \frac{\Delta\nu_{clock}}{\nu_{clock}} \approx -\sqrt\frac{\pi}{2} U \frac{d\Delta\alpha'_{E1}}{d\nu}\tilde S_0\sigma_{ASE} (\nu_{ASE}-\nu_{magic})
\end{equation}
With this expression, the fractional light shift can be estimated based on spectral measurements of the ASE profile. Notably, the shift scales as the product of the amplitude, width, and peak detuning of the ASE spectrum. For general non-Gaussian background spectra, the fractional shift can instead be evaluated numerically.
\\
\\
To validate our simple model given by equation (\ref{eq:fractional_shift}), we undertook a series of shift measurements while varying the amplifier current and temperature in an Yb optical lattice clock \cite{mcgrew2018atomic}. These measurements interleaved (i) generating the optical lattice with the TA versus (ii) an approximately background-free Ti:S at the same laser frequency, the latter of which is filtered by a volume Bragg grating bandpass filter with a 26 GHz bandwidth. The lattice intensity was controlled with feedback to acousto-optic modulators for each laser using a common photodetector signal and control electronics. The TA was phase-locked to the Ti:S, which itself was referenced to a stable optical cavity. A liquid crystal waveplate and polarizer was used to select between the two lasers, and the inactive beam was further attenuated by switching off the rf drive from an acousto-optic modulator in that beam's path. The active beam was coupled into a single optical fiber and delivered to the atomic system, passing through a polarizing beam splitter to provide linearly polarized light for the trap. This configuration ensured that the optical lattices formed by the two lasers had equal frequency, intensity, polarization, and spatial geometry. Furthermore, atomic cold collision shifts, which can substantially complicate lattice shift characterization \cite{nemitz2019modeling}, were suppressed in common mode due to the identical lattice conditions in the two cases. The experimental setup is shown in Fig. \ref{fig:experiments}a, except with the volume Bragg grating in the TA beam path removed.

\begin{figure*}[ht!]
\includegraphics[width=17.2cm]{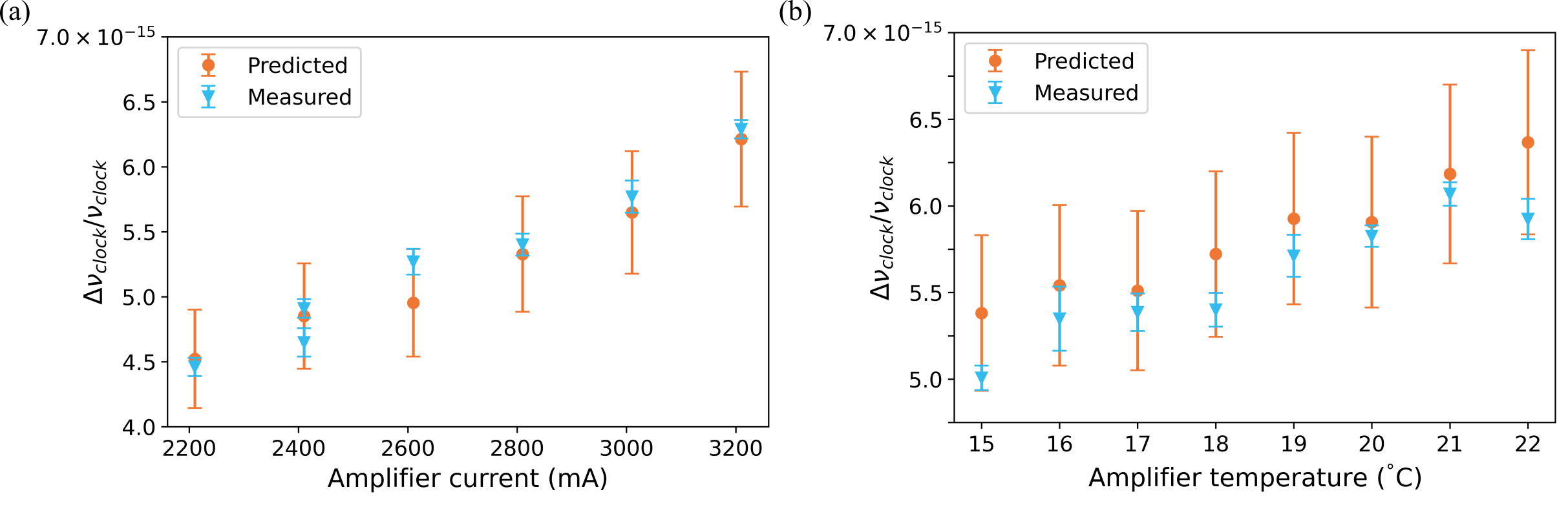}
\caption{\label{fig:unfiltered_measurements}Comparison of measured light shifts with predictions based on spectral characterization while varying a) amplifier current and b) amplifier temperature. The error bars represent 1$\sigma$ uncertainties given by the Allan deviation of the frequency difference between the TA and Ti:S. Uncertainties in predicted points are computed with standard propagation of error techniques based on uncertainties in the Gaussian model parameters, differential polarizability, and an assumed 5\% uncertainty in trap depth. }
\label{fig:manmade}
\end{figure*}

At each amplifier current and temperature, we additionally measured the ASE profile with an optical spectrum analyzer and extracted characteristic parameters with a Gaussian fit. Through motional sideband spectroscopy, the trap depth was assessed to be $U=53$. In calculating the shift, the trap depth was also scaled by an experimentally-determined factor of 0.71 to account for thermal averaging of the laser intensity by the motional distribution of atoms within the lattice \cite{brown2017hyperpolarizability}. The background light shift was predicted from (\ref{eq:shift_approx}) with an uncertainty given by standard propagation of error. The predictions are in good agreement with the measured shifts (Fig. \ref{fig:unfiltered_measurements}), demonstrating approximately linear dependence of the shift with amplifier current and temperature with slopes of $1.8\times 10^{-15}/$A and $1.6\times 10^{-16}/$K respectively. Therefore, to control the shift at the $10^{-18}$ level, the amplifier current and temperature would need to be stable at the 600 $\mu$A and 6 mK level.

Although this degree of control is practically realizable, previous efforts to characterize and control background light shifts have noted large time-varying effects \cite{le2013experimental, bloom2014building}. We observed substantial dependence of the shift on the alignment into the fiber used to deliver light to the atoms (NKT LMA-PM-15 large-mode-area photonic crystal fiber). Measurements of the ASE spectrum while intentionally degrading the alignment showed that significant spectral distortion could arise, which we attribute to a wavelength-dependent fiber coupling due to the spatial mismatch of the ASE and the fiber. This effect is discussed in greater detail in Section \ref{sec:uncertainty}-D.

For the optical clock measurements in Fig. \ref{fig:unfiltered_measurements}, the fiber pickoff for OSA measurements was located near the atoms, ensuring that any distortion induced by the photonic crystal fiber would be accounted for in the shift prediction. A multi-mode fiber was used to couple light to the OSA, and we verified that the observed laser spectrum was substantially less sensitive to misalignment than a single-mode alternative with 4.5 $\mu$m core diameter.

However, while these efforts make it possible to estimate the resulting background light shift at a given instant, degradation in fiber coupling at the few percent level modifies the shift by greater than $10^{-16}$. In contrast, control of the amplifier-dependent shifts at the $10^{-16}$ level requires only modest efforts to control the temperature and current. Therefore, we find that achieving robust control of background light shifts is primarily limited by beam pointing stability and virtually impossible at the $10^{-18}$ level. Instead, these shifts must be suppressed, such as realized with spectral filtering below.

\section{Spectral filtering\label{sec:filtering}}
Background light shifts can be reduced with optical bandpass filters centered at the magic wavelength. Various techniques have achieved passbands with $<$1 nm bandwidth, including complex prism geometries \cite{bakker1999new}, atomic vapor filters \cite{marling1979ultrahigh}, and various types of reflection gratings \cite{magnusson1992new, glebov2012volume} (potentially in conjunction with Fabry-Perot etalons \cite{lumeau2006tunable}). Due to their relative simplicity and compact size, we have chosen a volume Bragg grating. Near normal incidence, the reflection transfer function is approximately \cite{hellstrom2007finite}
\begin{equation}\label{eq:transfer-function}
    R(\nu)=\frac{\sinh^2\phi(\nu)}{\cosh^2\phi(\nu)-\Gamma(\nu)^2}
\end{equation}
where 
\begin{equation}\label{eq:transfer-arg}
    \phi(\nu) = \frac{2 \pi d\nu}{c}\sqrt{ \frac{n_1^2}{4} - \bigg (\frac{\nu_0}{\nu}-1\bigg )^2}
\end{equation}
and 
\begin{equation}
    \Gamma(\nu) = \frac{2}{n_1}\bigg (\frac{\nu_0}{\nu}-1\bigg )
\end{equation}
Here, $n_1$ is the index modulation depth and $\nu_0$ is the center frequency. Two distinct behaviors are evident in this transfer function. Near the center frequency ($\nu\approx\nu_0)$, the transfer function is approximately constant and reaches a peak reflectivity
\begin{equation}\label{eq:peak-reflectivity}
    R(\nu_0) = \tanh^2 \frac{\pi d n_1\nu_0}{c}
\end{equation}
Far from the design frequency, the transfer function asymptotically varies as 
\begin{equation}\label{eq:asymptotic_vbg}
    R(\nu)\approx \frac{n_1^2}{4}\frac{\sin^2[\frac{\pi d}{c}(\nu-\nu_0)]}{(\frac{\nu_0}{\nu}-1 )^2},
\end{equation}
which is the product of a periodic modulation and a second-order filter rolloff. The transition point between the pass and rejection bands occurs where (\ref{eq:transfer-arg}) changes from real to imaginary, at $\nu=2\nu_0/(2\pm n_1)$. The bandwidth can thus be approximated as $\Delta\nu_{BW}\approx\nu_0 n_1$. Representative values for gratings used in this work are $n_1=2.3\times 10^{-5}$ and $d=22$ mm, resulting in a theoretical bandwidth of 9.2 GHz and a peak reflectivity of 94\%. 

\begin{figure}[t!]
\includegraphics[width=8.6cm]{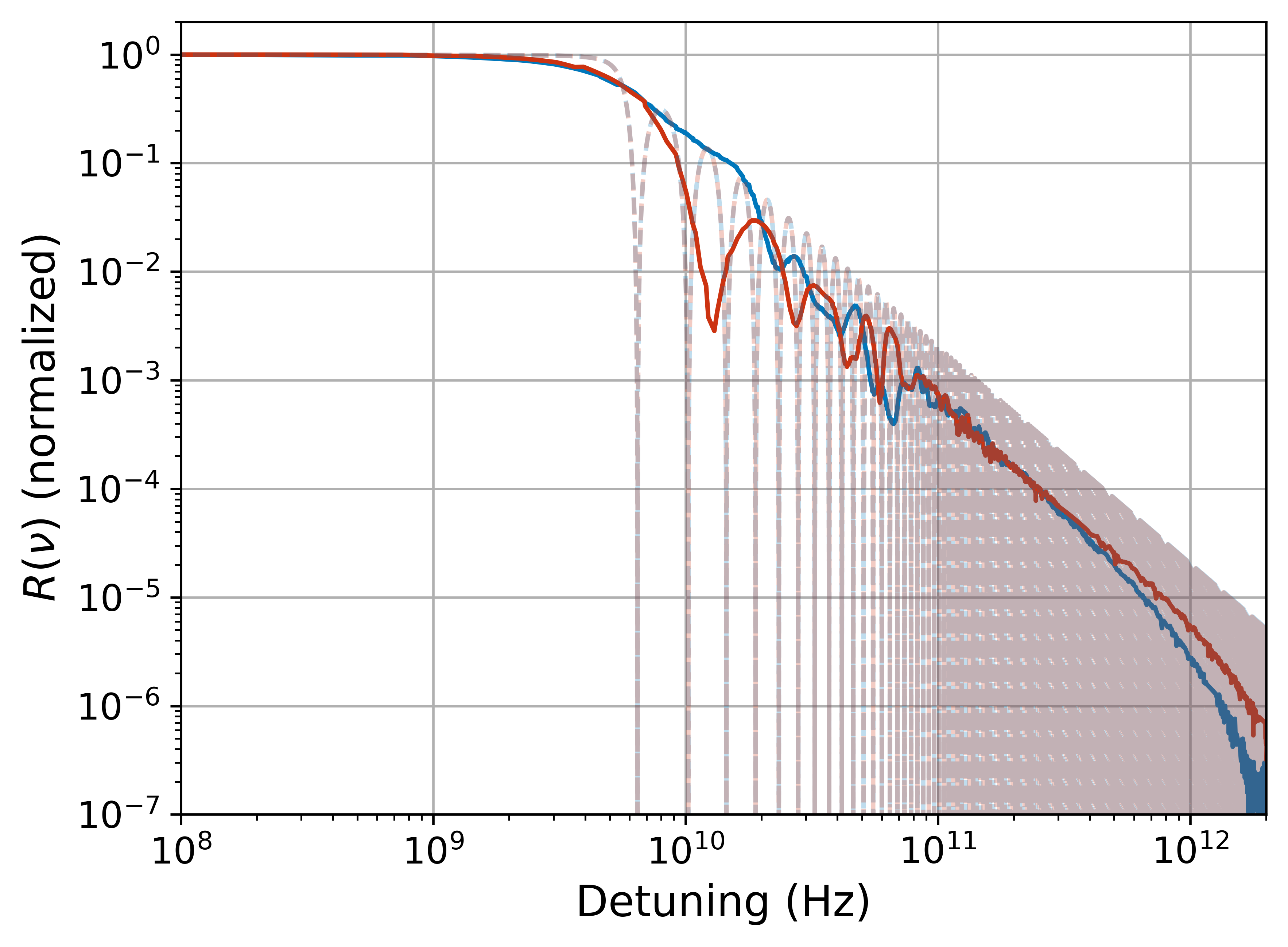}
\caption{\label{fig:transfer_function}Volume Bragg grating transfer function, measured by varying the Ti:S frequency and measuring the reflected power. Positive detuning is shown in blue and negative in red. The theoretical profile is shown in light dashed lines.}
\end{figure}

Volume Bragg gratings induce a wavelength-dependent distortion of the spatial beam profile \cite{hellstrom2007finite}. To avoid complicated spatially-varying shifts in an optical lattice, we clean the beam profile with a single-mode optical fiber between the grating and the atoms. The actual transfer function is therefore the product of the spectral selectivity of the grating with the distortion-induced coupling losses of the fiber. This combined transfer function was experimentally characterized by measuring the reflected and subsequently fiber-coupled power from a Ti:S beam incident on the grating while varying the Ti:S frequency over several THz (Fig. \ref{fig:transfer_function}). By adjusting the collimation of the incident beam, a peak in-band reflectivity of 98\% was achieved. The observed transfer function has a 20 dB/decade rolloff outside of an 11.5 GHz passband. The asymmetric ripples arise due to fiber coupling losses from beam distortion, while the characteristic modulation given by the idealized numerator of (\ref{eq:asymptotic_vbg}) is absent.
\\
\\
The light shift from the filtered background spectrum can be estimated by applying the filter transfer function (\ref{eq:transfer-function}) to the ASE profile (\ref{eq:gaussian-profile}) and numerically integrating the shift spectral density (\ref{eq:fractional_shift}); we treat the experimentally-observed departures from the theoretical transfer function as an uncertainty in Section \ref{sec:uncertainty}. For typical operating parameters of 3210 mA amplifier current and 20 $^{\circ}$C temperature, the predicted shift is $8\times 10^{-21}$ for the experimental trap depth $U=53$, including the thermal averaging factor of 0.71. We note that another recent work has estimated ASE light shifts below the $10^{-19}$ level using a VBG bandpass filter with 35 pm bandwidth in combination with a cavity with 6 MHz linewidth \cite{takamoto2020test}. In either case, as discussed below, these calculations rely on a number of assumptions that must be verified to ensure reliable shift suppression at the predicted levels.

\begin{figure*}[t!]
\includegraphics[width=17.2cm]{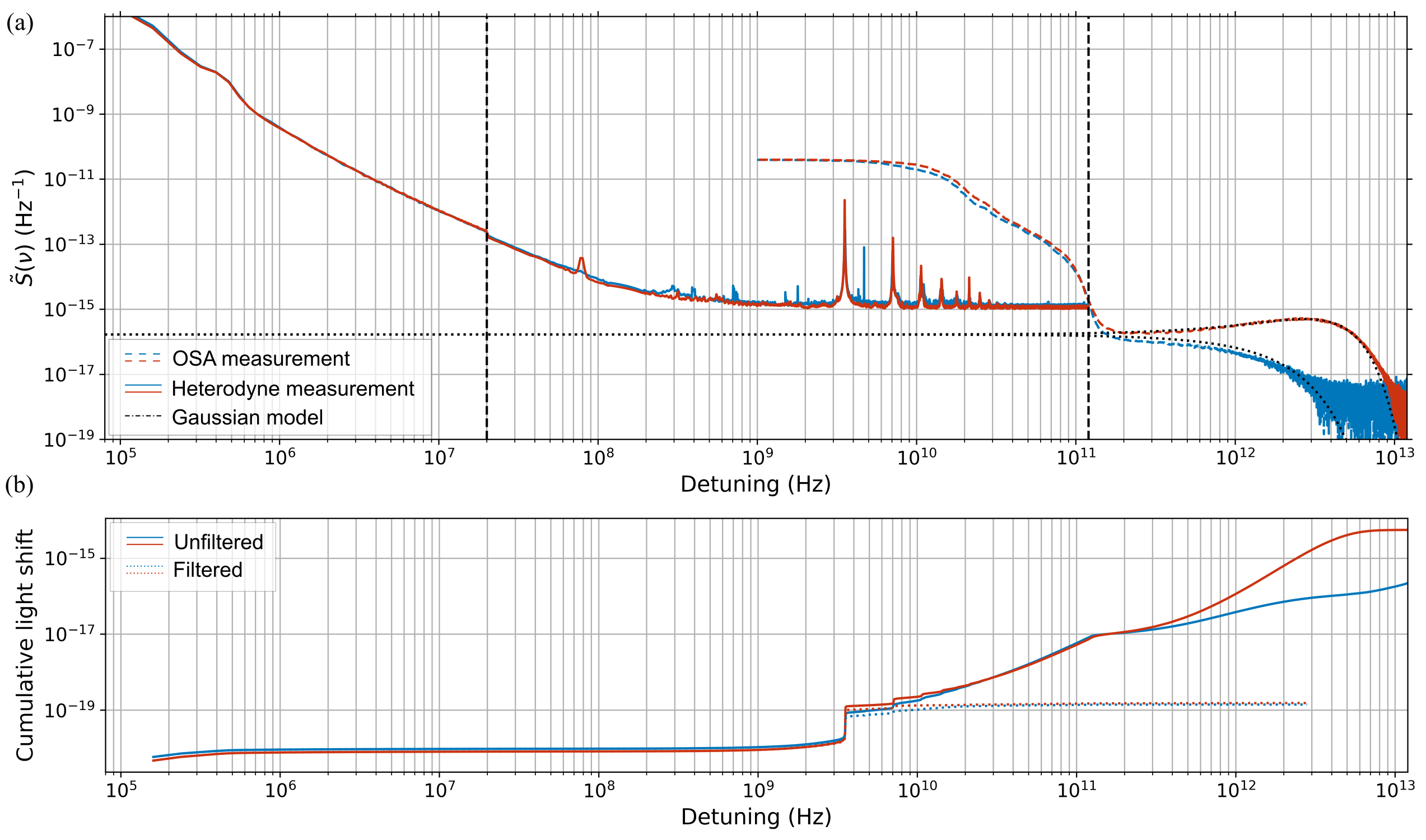}
\caption{\label{fig:composite_spectra}a) Composite spectral measurement of the tapered amplifier system, obtained from three different measurements: a $\pm$20 MHz heterodyne measurement with 200 kHz resolution bandwidth (RBW), a $\pm100$ GHz heterodyne measurement with 5 MHz RBW, and a $\pm10$ nm optical spectrum analyzer measurement with 20 pm RBW. Black dashed lines mark the boundaries of the three measurements. The power spectral density normalized to the carrier power is plotted against the absolute value of the detuning from the magic wavelength carrier, with blue (red) showing spectral content at higher (lower) frequencies than the magic wavelength. The dash-dotted line shows a Gaussian fit to the ASE profile measured by the OSA.
b) Cumulative shift upper bound, calculated by integrating the shift spectral density from the composite spectrum from the magic wavelength to a variable frequency. The filtered shift spectral density is calculated using the measured VBG transfer function in Fig. \ref{fig:transfer_function}.
}
\end{figure*}

\section{\label{sec:uncertainty} Shift estimate uncertainty}
While the theoretical framework outlined in Sections \ref{sec:shifts} and \ref{sec:filtering} predicts shifts that can be made negligible compared to the uncertainty budgets of current state-of-the-art clocks, it relies on several empirically-motivated assumptions and is potentially vulnerable to fluctuating systematic effects. In this section, we analyze these assumptions and effects to determine a conservative upper bound on the background light shift during normal operating conditions. All uncertainties are presented for the experimental trap depth $U=53$ as described in Section \ref{sec:shifts}.

\subsection{Near-carrier spectrum}
Within 100 GHz of the magic wavelength, our ability to resolve the spectrum is limited by the filter shape and optical rejection ratio of the optical spectrum analyzer. This limitation is referred to as the "close-in dynamic range," which is specified to be 60 dB at $\pm 0.5$ nm from a spectral peak for the OSA used in these measurements. This prevents direct observation of the background spectrum, potentially concealing strong spectral features associated with deviations from the Gaussian ASE profile or other lasing modes. Additionally, it could obscure any interaction between the coherent lasing carrier and the ASE background, which would otherwise challenge the assumption that they can be treated separately. These phenomena are especially worrying due to their proximity to the VBG filter passband, as any shifts in this region may persist even after filtering.

To highlight the need for caution here, we calculate the expected light shift from one side of the OSA spectrum (red detuning) observed in Fig. 1 inset, rather than using the Gaussian model.  For the case of the Ti:S laser, without the VBG filter, we find a shift estimate of $4\times 10^{-15}$.  After application of the filter, the remaining shift is still $6\times 10^{-16}$.  Spectral characterizations using the OSA are therefore insufficient to confidently constrain the background light shifts below the $10^{-18}$ level, even for the spectrally pure case of the Ti:S laser.

To improve the spectral characterization close to the carrier and thus tighten these constraints, we made an optical heterodyne measurement between the TA and Ti:S. The two beams were overlapped and directed onto a fiber-coupled detector and the beat signal was measured with an RF signal analyzer. Measurements were made in the heterodyne frequency range of 250 MHz to 4 GHz, while the Ti:S frequency was stepped in several GHz intervals to facilitate measurement out to $\pm100$ GHz offset from the TA carrier. Spurious spectral components corresponding to the complex-conjugate Fourier ambiguity were removed using an image-rejection algorithm. For each nominal detuning in the scan, the relative frequency was varied by approximately $\pm 100$ MHz while several traces were recorded. When these traces were overlaid on a common frequency axis, the ``real'' signal overlapped, while the negative-frequency components varied in apparent frequency, allowing them to be identified and removed. As the heterodyne measurements are unable to differentiate between broadband features such as ASE at positive and negative heterodyne frequencies, we note that the ASE power is conservatively overestimated by 6 dB in the range below 100 GHz detuning.

A number of noise processes potentially contribute to the heterodyne measurement sensitivity. Thermal fluctuations (Johnson-Nyquist noise) yield a noise power $k_BTB$ for a measurement bandwidth $B$, resulting in a noise floor of $-114$ dBm/MHz at room temperature. Additionally, our electronic signal analyzer has a noise floor at -95 dBm/MHz. Another significant noise source was photoelectric shot noise. Consider a photodiode with responsivity $\mathcal R$, terminating resistance $R$, and an additional power gain of $G$. In terms of the optical powers of each beam, $P_{TA}$ and $P_{Ti:S}$, the shot noise is  $\delta P_e=2e\mathcal R (P_{TA}+P_{Ti:S})RGB$, while the signal power is $P_e=2\mathcal R^2 P_{TA}P_{Ti:S} RG$. The resulting signal-to-noise ratio is
\begin{equation}
    \frac{P_e}{\delta P_e} = \frac{\mathcal R}{eB} \frac{P_{TA}P_{Ti:S}}{P_{TA}+P_{Ti:S}}
\end{equation}
Since we measured the electronic power spectral density, rather than the total electronic power, the relevant bandwidth in this calculation is the resolution bandwidth of the signal analyzer, which was chosen to be several orders of magnitude smaller than the photodiode bandwidth. As long as shot-noise-limited performance is achieved in spite of other noise processes in the measurement system, the photodiode bandwidth only determines the spectral span accessible at a single choice of Ti:S laser detuning; therefore, the product of the responsivity and the linearity threshold is the figure of merit for the photodiode. For these measurements, we used an InGaAs photodiode (Discovery Semiconductors xHLPD) with 0.25 A/W responsivity, 4 GHz bandwidth, and a linearity threshold of around 100 mW, achieving a shot noise floor of -94 dBm/MHz. 

In Figure \ref{fig:composite_spectra}a, we combine OSA and heterodyne data to show the spectrum over eight decades of frequency range, from hundreds of kHz to tens of THz offset from the magic wavelength for $^{171}$Yb. We note that, although the heterodyne measurement is able to constrain background light much better than the OSA (by up to five orders of magnitude for detunings $<$100 GHz), shot noise still prevents direct observation of the ASE profile in this region as interpolated by the Gaussian fit \footnote{Because the heterodyne measurements include ASE spectrum from both positive and negative heterodyne frequencies, essentially doubling the expected measurement of the ASE amplitude, comparison of the observed spectrum in this frequency region should be made relative to the interpolated Gaussian fit plus an additional 6 dB.}.

Using equation \ref{eq:fractional_shift}, we compute an upper bound on the shift spectral density, and determine the cumulative shift by integrating the shift spectral density outwards from the magic wavelength, shown in Fig. \ref{fig:composite_spectra}b. The VBG filter limits the region where the shift spectral density is large enough to cause appreciable shifts, causing convergence of the cumulative shift near the edge of the passband. Our upper bounds on the shift established by this composite spectrum are $6\times 10^{-15}$ without the filter and $1.4\times 10^{-19}$ filtered. In the filtered case, a notable limitation is the presence of spectral features detuned from the carrier at multiples of 3.5 GHz, corresponding to other longitudinal modes of the seed laser. While there is partial cancellation of red- and blue-detuned features, we use the one-sided estimate as an upper bound, noting that the shift resulting from these peaks is around the $10^{-19}$ level after filtering. Similar features were observed on the Ti:S spectrum, resulting in one-sided shifts at the $10^{-20}$ level. The dominance of these features relative to the overall constraints highlights the importance of careful spectral characterization within the filter passband, which is beyond the limits of typical optical spectrum analyzers. Even without a separate laser for heterodyne characterization, monitoring the self-beating near-carrier spectrum on a high-bandwidth photodiode may be sufficient to detect multi-mode operation at the level required for $10^{-19}$ uncertainty.

\subsection{Model extrapolation error}
The range over which the ASE profile is observable is limited to approximately $\pm10$ THz by the noise floor of the OSA (-80 dBm). The assumption of Gaussian ASE profiles is empirically supported, but extrapolation below the noise floor of the OSA is questionable, as deviations from the assumed Gaussian profile or the observed VBG transfer function could modify the shift. To determine a conservative upper bound on the shift due to spectral content below the OSA noise floor, we assume a constant power spectral density at the noise floor and integrate from the magic wavelength to the edge of our mirror transmission window at 273 THz, again neglecting partial cancellation from the opposite detuning. We additionally assume that the VBG transfer function is constant at -70 dB within the range, neglecting increased attenuation below the noise floor of Fig. (\ref{fig:transfer_function}). This yields a constraint of $-8\times 10^{-19}$ for the unfiltered spectrum and $-2\times 10^{-20}$ with the filter applied. These constraints could be further tightened by employing a multipass Bragg filter \cite{ott2015high} to more strongly attenuate light outside the passband, or by improving the signal-to-noise ratio of the characterizations of the laser spectrum or VBG transfer function.

We also observe that the Gaussian model falls off slightly faster than the observed ASE profile far from the peak (Fig. \ref{fig:osa-measurements}). By directly integrating the observed profile, we determine that this contributes an error of $<5\%$ to the unfiltered shift estimate. This model error therefore contributes negligibly when the bandpass filter is employed.

\subsection{Grating transfer function}
For the purpose of simple numerical evaluation of the shift, the analytical bandpass transfer function (\ref{eq:transfer-function}) was used. However, realistic optical bandpass filters can diverge from the theoretical plane-wave transfer function (\ref{eq:transfer-function}) due to the effects of finite beams \cite{hellstrom2007finite}, non-monochromatic Gaussian beams with imperfect collimation \cite{ciapurin2005modeling} or potentially the non-Gaussian beam profiles characteristic of tapered amplifiers. We note that the filtered shift has low sensitivity to errors in the determination of the filter bandwidth - doubling the bandwidth would still produce a shift of only $-3\times 10^{-20}$. We take this value as an upper bound associated with the grating transfer function. Additionally, the experimentally-characterized reflectivity was used to compute the upper bound in Section \ref{sec:uncertainty}-A, accounting for deviations from the theoretical profile.

Because the center wavelength of a Bragg filter is tunable through the angle of incidence of light onto the grating, angular misalignment due to improper setup or beam pointing drift results in a wavelength offset of the passband. We estimate the angular bandwidth of our grating to be $200\text{ }\mu$rad \cite{hellstrom2007finite}. By calculating the shift for varying center frequency, we find a dependence of $8\times 10^{-21}$ per GHz. Therefore, detuning to the edge of the passband modifies the shift by $4\times 10^{-20}$. We take this as a conservative uncertainty on the shift due to filter misalignment.

\subsection{Fiber misalignment}
During the unfiltered clock shift measurements in Section \ref{sec:shifts}, it was observed that small changes in fiber alignment could change the measured shift substantially. To assess the magnitude of this effect, we experimentally degraded the fiber alignment using a mirror mount before the fiber delivering light to the atoms as well as the pickoff for OSA spectral characterization. We use the relative decrease in transmitted power to parametrize the misalignment, with an additional sign factor corresponding to the direction of actuation of the mirror mount. Using OSA spectra measured at varying misalignment, we compute the background light shift with and without the filter. The shifts and OSA spectra as a function of misalignment are presented in Fig. \ref{fig:seed-attenuation}a.

Without filtering, the shift is susceptible to fiber misalignment effects at the level of $3\times 10^{-16}$ per percent of transmitted power loss. With filtering, this dependence is suppressed to $4\times 10^{-23}$ per percent. Therefore, beam pointing instability contributes negligibly to the uncertainty budget with an assumed worst-case degradation of 10\%.

\subsection{Seed-amplifier coupling}
\begin{figure}[h!]
\includegraphics[width=8.6cm]{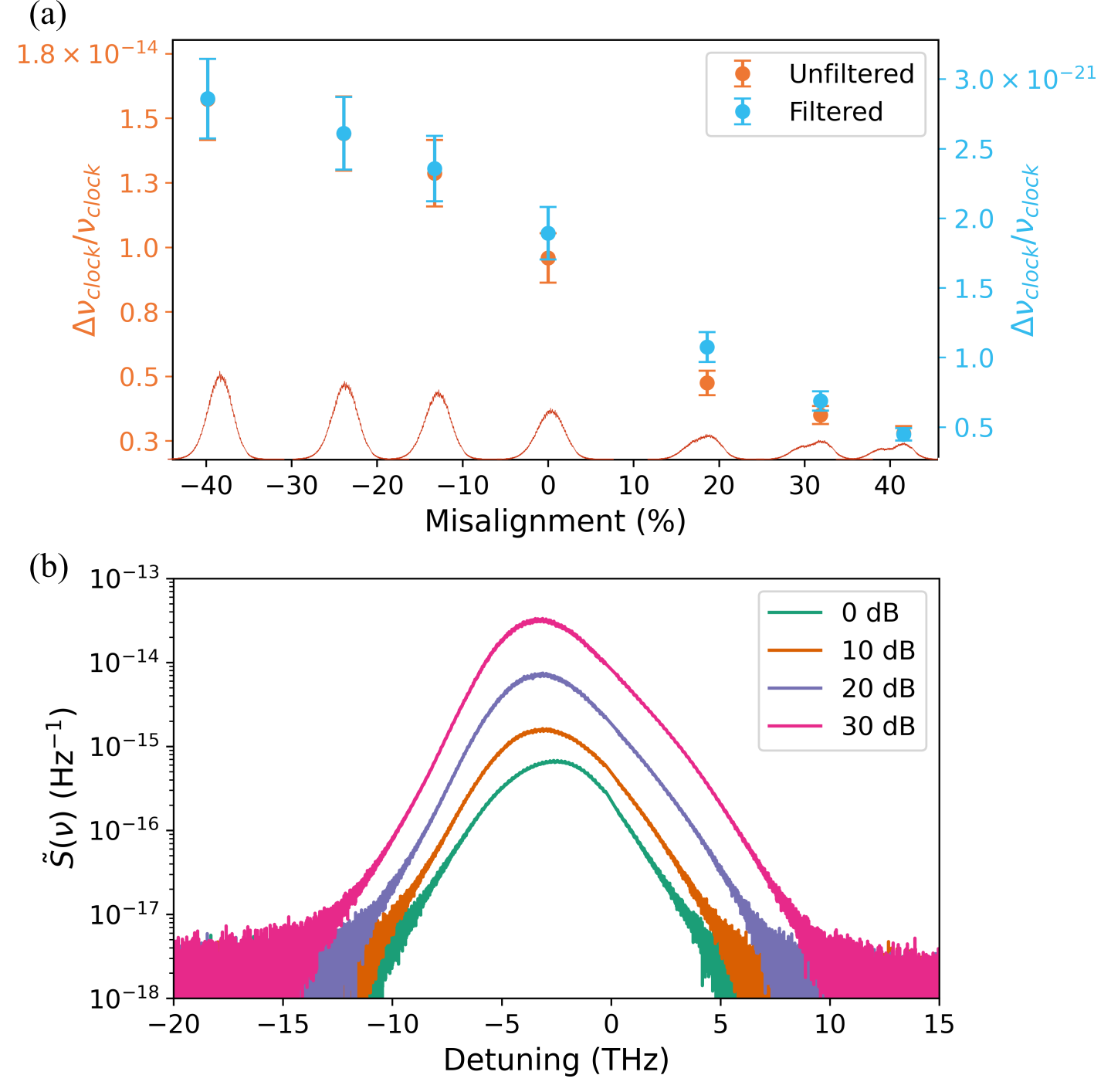}
\caption{\label{fig:seed-attenuation} a) ASE spectra (red) and estimates of the associated shifts with (red) and without (orange) spectral filtering as a function of misalignment. b) ASE spectra at varying seed attenuation relative to a nominal seed power of 35 mW. The spectra are processed to remove the carrier and interpolate the missing ASE spectrum.}
\end{figure}

The shift could also potentially change with increase in relative ASE level due to a degradation of seed-amplifier coupling. To characterize the sensitivity of this effect, we attenuated the seed laser with varying neutral density filters and measured the resulting optical spectrum with an OSA (Fig. \ref{fig:seed-attenuation}b). With 30 dB of seed attenuation, the ASE level is increased only by a factor of approximately 60. Noting that seed-amplifier misalignment would result in a decrease in output power which would be noticed and quickly corrected, we very conservatively assume a maximum seed attenuation of 10 dB, producing a shift of $-3\times 10^{-20}$.

\subsection{Total estimated shift uncertainty}
Our uncertainty budget encompasses several constraints based on characterization limitations as well as uncertainty estimates for several time-varying systematics. To determine the total uncertainty, we model the former case as uniform distributions truncated at our upper bound and accordingly compute a standard deviation. The latter effects are assumed to stem from Gaussian distributions with $2\sigma$ widths given by our worst-case estimates. We report all uncertainties at the $1\sigma$ level in Table \ref{tab:budget}. The total uncertainty on the magnitude of the filtered background light shift, computed as the quadrature sum of individual effects, is $5\times 10^{-20}$, most significantly limited by our ability to constrain the amplitude of residual spectral content near the passband. We note that, due to the strong out-of-band rejection of the filter, even highly conservative constraints on spectral content below the OSA noise floor at large detunings exceeding 10 THz do not contribute substantially to our overall constraints. 
\begin{table}[]
\label{tab:budget}
\caption{Background light shift uncertainty budget}

\begin{tabular}{|l|l|}
\hline
\textbf{Effect}           & \textbf{Uncertainty ($\times 10^{-20}$)} \\ \hline
Near-carrier spectrum     & 4                                \\ \hline
Model extrapolation error      & 0.6                                \\ \hline
Grating transfer function & 0.8                                \\ \hline
Grating misalignment      & 2                                \\ \hline
Fiber misalignment        & 0.02                                \\ \hline
Seed-amplifier coupling   & 2                                \\ \hline
\textbf{Total}            & 5                               \\ \hline
\end{tabular}
\end{table}

\section{Direct experimental validation\label{sec:experimental_validation}}
\begin{figure*}[ht!]
\centering
\includegraphics[width=17.2cm]{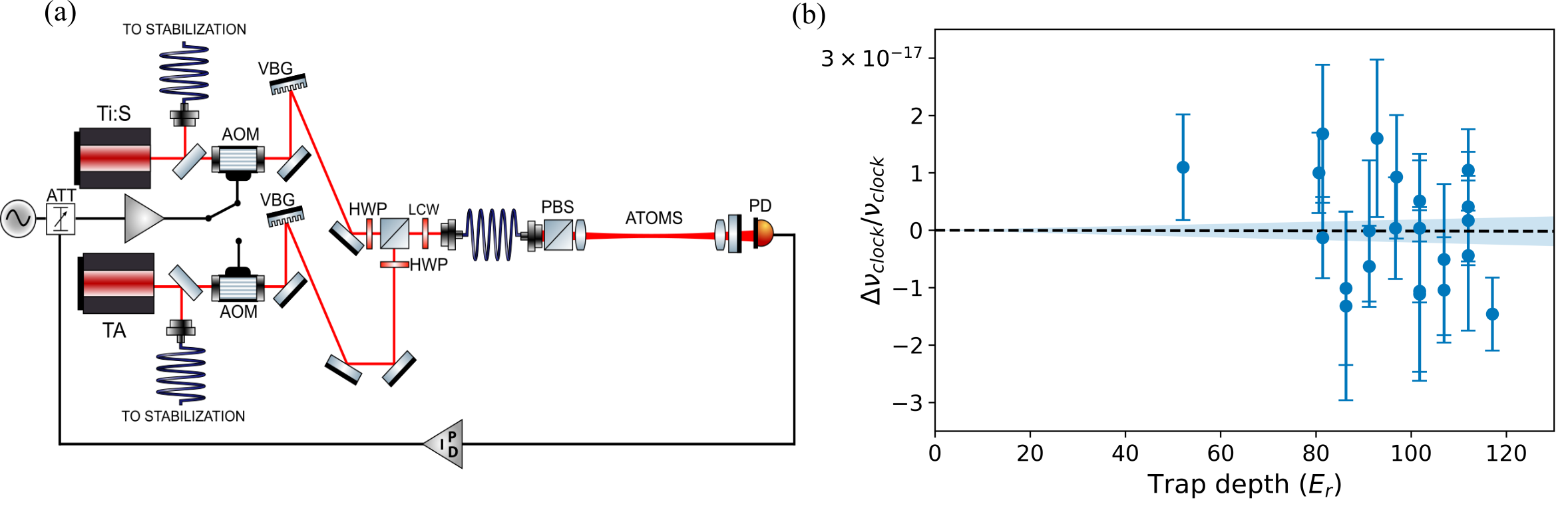}
\caption{\label{fig:experiments} \textbf{a)} Experimental setup for interleaved clock comparisons. During each cycle, the active beam is selected using a liquid crystal waveplate (LCW) and polarization beamsplitter (PBS). The intensity of the active beam, monitored with a photodetector (PD) behind the lattice retroreflector, is stabilized through the rf drive power of an acousto-optic modulator (AOM) using a variable attenuator (ATT); an rf switch selecting the active AOM further attenuates the inactive beam. Each laser is filtered by a volume Bragg grating (VBG); the Ti:S filter has 26 GHz bandwidth, while the TA filter has 11.5 GHz bandwidth. In the unfiltered measurements in Section \ref{sec:shifts}, the volume Bragg grating in the TA beam path was replaced with a mirror. \textbf{b)} Fractional clock shifts between the filtered TA and reference Ti:S configurations. The blue band shows the $1\sigma$ confidence region on the linear fit.}
\end{figure*}

To corroborate our upper bounds on the background light shift, we measured the filtered shift using the interleaved atomic clock configuration from Section \ref{sec:shifts} with the addition of the volume Bragg grating filtering the TA (Fig. \ref{fig:experiments}a). This allows us to compare the background light shift from the filtered TA laser system relative to the intrinsically more spectrally pure filtered Ti:S laser system. At an operational amplifier current of 3210 mA and a temperature of 20 $^{\circ}$C, the frequency of the Yb clock transition was measured while switching between the filtered Ti:S and filtered TA for a number of trap depths between 50-120 $E_r$ (Fig. \ref{fig:experiments}b). Each measurement was terminated at a statistical uncertainty near $1\times 10^{-17}$ as given by the Allan deviation extrapolated to the full measurement time. As before, the influence of cold collision shifts on this measurement were made negligible by operating at a constant, low atom number density. A weighted linear regression with zero intercept was used to compare data at different trap depths and scale to an operational trap depth of 50 $E_r$, at which we evaluate the shift to be $(0.1\pm 1.0)\times 10^{-18}$. The reduced $\chi^2$ of the fit is 1.02, indicating that the scatter of the various measurements are consistent with their stated uncertainties. Thus, under typical operating conditions, we find that the background light shift of the filtered TA system is consistent with zero at the $1\times 10^{-18}$ level.

\section{Conclusion}
The background spectra of amplified diode lasers can cause Hz-level or higher shifts of the clock transition in optical lattice clocks, and therefore these systems are frequently avoided in favor of spectrally-pure alternatives like the Ti:sapphire. In this work, we have characterized the background of an amplified diode laser and estimated that the shift can be suppressed from $>10^{-15}$ to $8\times 10^{-21}$ for typical experimental conditions with volume Bragg grating bandpass filter. We have assessed several systematic effects that could either skew our estimates or cause the shift to vary with time, constraining the shift to be smaller than $5\times 10^{-20}$. This constraint was corroborated with a direct atomic measurement in an Yb optical lattice clock, demonstrating a shift consistent with zero at the $1\times 10^{-18}$ level. This shift characterization supports use of amplified diode lasers within the uncertainty budgets of state-of-the-art optical lattice clocks, offering an alternative to Ti:S systems with lower size, weight, power, and cost. Furthermore, the methods presented are general for characterizing background spectra in diverse laser sources, relevant to applications in precision optical interferometry such as molecular spectroscopy \cite{kondov2019molecular}, atom interferometry \cite{zhan2015investigating}, quantum computing \cite{dumke2002micro}, and gravitational wave detection \cite{shortt2019lisa}.
\\
\\
We appreciate useful discussions and technical support from F. Quinlan, K. Beloy, C. Nelson, A. Hati and T. Fortier. This work was supported by NIST, DARPA, and NSF QLCI Award OMA - 2016244. Contributions to this article by workers at NIST, an agency of the U.S. Government, are not subject to U.S. copyright.

\appendix 
\section{\label{sec:incoherence}ASE spatial averaging}
\begin{figure}[h!]
\includegraphics[width=8.6cm]{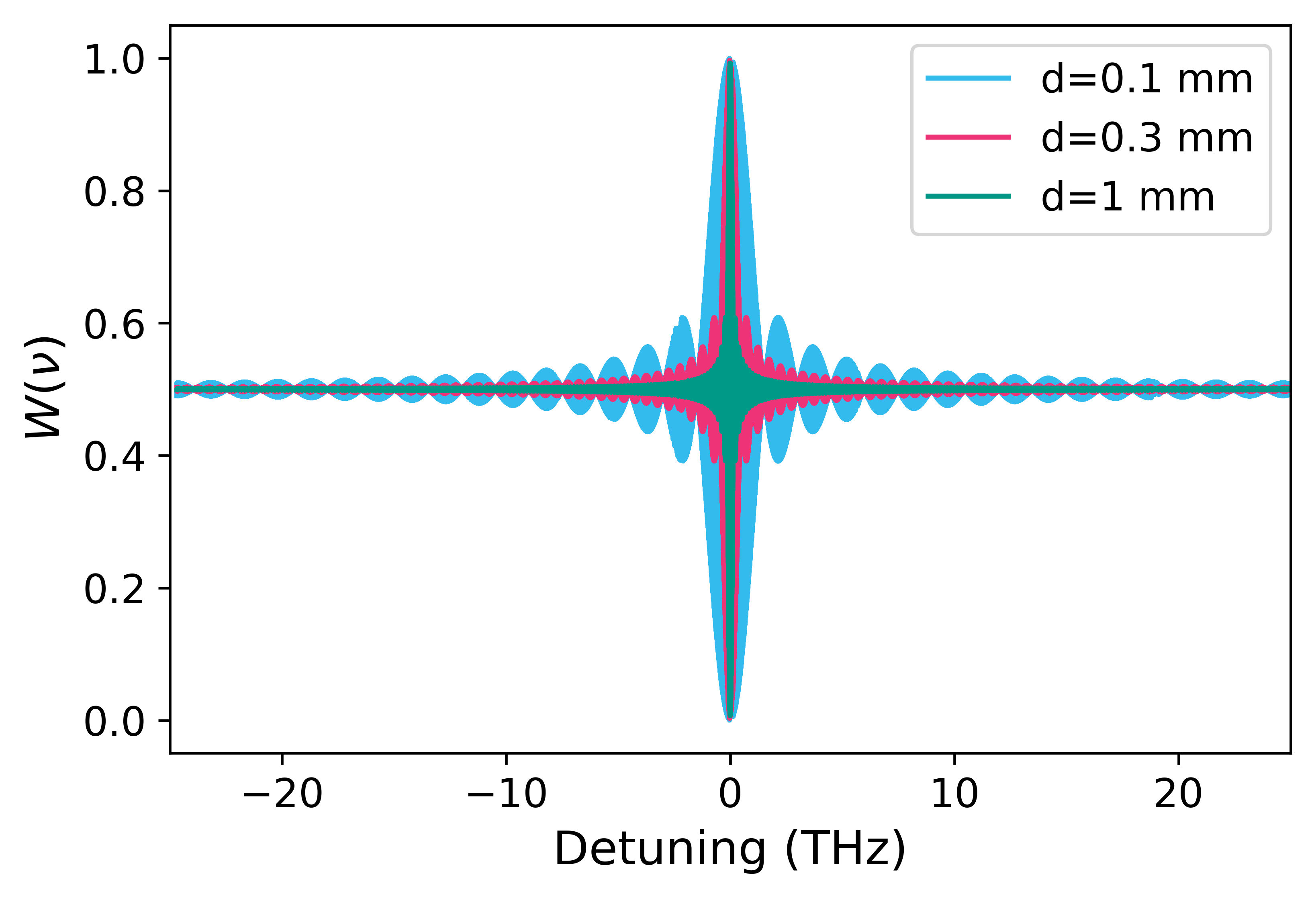}
\caption{\label{fig:weight_factor} Weighting function due to ASE incoherence relative to the trap, shown for several different characteristic trap sizes and a distance $D=1$ m.}
\end{figure}

After reflecting from the lattice retroreflector, the ASE forms a continuum of standing waves that are generally incoherent with the trap. Therefore, the average ASE intensity experienced by trapped atoms will be reduced. We estimate this effect by assuming that atoms are trapped in discrete positions $x_n=n\lambda_{magic}/2$ over a spatial distance $d$. The ASE intensity at a wavelength $\lambda$ is therefore weighted by a factor
\begin{equation}
    W(\lambda)=\frac{1}{N}\sum_{n=-N/2}^{N/2}\cos^2 \bigg (\frac{\pi n\lambda_{magic}}{\lambda}+\phi(\lambda) \bigg )
\end{equation}
where $N=2d/\lambda_{magic}$ and 
\begin{equation}
    \phi(\lambda)=2\pi\bigg (\frac{D}{\lambda}-\frac{D}{\lambda_{magic}}\bigg )
\end{equation}
is the differential phase accumulated between the trap and ASE at $x=0$. The round-trip distance between the center of the optical lattice and the retroreflector is given by $D$. In general, the weighting factor oscillates around 1/2 with a period and decay rate depending on the size of the trap as shown in Fig. \ref{fig:weight_factor}. At the magic wavelength, $W(\lambda_{magic})=1$, and thus equation (\ref{eq:E1_shift}) reduces to (\ref{eq:trap_shift}) for a monochromatic background spectrum at the magic wavelength, $S(\nu)=P\delta(\nu-\nu_{magic})$.

\begin{thebibliography}{44}%
\makeatletter
\providecommand \@ifxundefined [1]{%
 \@ifx{#1\undefined}
}%
\providecommand \@ifnum [1]{%
 \ifnum #1\expandafter \@firstoftwo
 \else \expandafter \@secondoftwo
 \fi
}%
\providecommand \@ifx [1]{%
 \ifx #1\expandafter \@firstoftwo
 \else \expandafter \@secondoftwo
 \fi
}%
\providecommand \natexlab [1]{#1}%
\providecommand \enquote  [1]{``#1''}%
\providecommand \bibnamefont  [1]{#1}%
\providecommand \bibfnamefont [1]{#1}%
\providecommand \citenamefont [1]{#1}%
\providecommand \href@noop [0]{\@secondoftwo}%
\providecommand \href [0]{\begingroup \@sanitize@url \@href}%
\providecommand \@href[1]{\@@startlink{#1}\@@href}%
\providecommand \@@href[1]{\endgroup#1\@@endlink}%
\providecommand \@sanitize@url [0]{\catcode `\\12\catcode `\$12\catcode
  `\&12\catcode `\#12\catcode `\^12\catcode `\_12\catcode `\%12\relax}%
\providecommand \@@startlink[1]{}%
\providecommand \@@endlink[0]{}%
\providecommand \url  [0]{\begingroup\@sanitize@url \@url }%
\providecommand \@url [1]{\endgroup\@href {#1}{\urlprefix }}%
\providecommand \urlprefix  [0]{URL }%
\providecommand \Eprint [0]{\href }%
\providecommand \doibase [0]{https://doi.org/}%
\providecommand \selectlanguage [0]{\@gobble}%
\providecommand \bibinfo  [0]{\@secondoftwo}%
\providecommand \bibfield  [0]{\@secondoftwo}%
\providecommand \translation [1]{[#1]}%
\providecommand \BibitemOpen [0]{}%
\providecommand \bibitemStop [0]{}%
\providecommand \bibitemNoStop [0]{.\EOS\space}%
\providecommand \EOS [0]{\spacefactor3000\relax}%
\providecommand \BibitemShut  [1]{\csname bibitem#1\endcsname}%
\let\auto@bib@innerbib\@empty
\bibitem [{\citenamefont {Katori}\ \emph {et~al.}(2003)\citenamefont {Katori},
  \citenamefont {Takamoto}, \citenamefont {Pal’{c}hikov},\ and\ \citenamefont
  {Ovsiannikov}}]{katori2003ultrastable}%
  \BibitemOpen
  \bibfield  {author} {\bibinfo {author} {\bibfnamefont {H.}~\bibnamefont
  {Katori}}, \bibinfo {author} {\bibfnamefont {M.}~\bibnamefont {Takamoto}},
  \bibinfo {author} {\bibfnamefont {V.}~\bibnamefont {Pal’{c}hikov}},\ and\
  \bibinfo {author} {\bibfnamefont {V.}~\bibnamefont {Ovsiannikov}},\
  }\bibfield  {title} {\bibinfo {title} {Ultrastable {O}ptical {C}lock with
  {N}eutral {A}toms in an {E}ngineered {L}ight {S}hift {T}rap},\ }\href@noop {}
  {\bibfield  {journal} {\bibinfo  {journal} {Physical {R}eview {L}etters}\
  }\textbf {\bibinfo {volume} {91}},\ \bibinfo {pages} {173005} (\bibinfo
  {year} {2003})}\BibitemShut {NoStop}%
\bibitem [{\citenamefont {Bloch}\ \emph {et~al.}(2008)\citenamefont {Bloch},
  \citenamefont {Dalibard},\ and\ \citenamefont {Zwerger}}]{bloch2008many}%
  \BibitemOpen
  \bibfield  {author} {\bibinfo {author} {\bibfnamefont {I.}~\bibnamefont
  {Bloch}}, \bibinfo {author} {\bibfnamefont {J.}~\bibnamefont {Dalibard}},\
  and\ \bibinfo {author} {\bibfnamefont {W.}~\bibnamefont {Zwerger}},\
  }\bibfield  {title} {\bibinfo {title} {Many-body physics with ultracold
  gases},\ }\href@noop {} {\bibfield  {journal} {\bibinfo  {journal} {Reviews
  of {M}odern {P}hysics}\ }\textbf {\bibinfo {volume} {80}},\ \bibinfo {pages}
  {885} (\bibinfo {year} {2008})}\BibitemShut {NoStop}%
\bibitem [{\citenamefont {Briegel}\ \emph {et~al.}(2000)\citenamefont
  {Briegel}, \citenamefont {Calarco}, \citenamefont {Jaksch}, \citenamefont
  {Cirac},\ and\ \citenamefont {Zoller}}]{briegel2000quantum}%
  \BibitemOpen
  \bibfield  {author} {\bibinfo {author} {\bibfnamefont {H.-J.}\ \bibnamefont
  {Briegel}}, \bibinfo {author} {\bibfnamefont {T.}~\bibnamefont {Calarco}},
  \bibinfo {author} {\bibfnamefont {D.}~\bibnamefont {Jaksch}}, \bibinfo
  {author} {\bibfnamefont {J.~I.}\ \bibnamefont {Cirac}},\ and\ \bibinfo
  {author} {\bibfnamefont {P.}~\bibnamefont {Zoller}},\ }\bibfield  {title}
  {\bibinfo {title} {Quantum computing with neutral atoms},\ }\href@noop {}
  {\bibfield  {journal} {\bibinfo  {journal} {Journal of {M}odern {O}ptics}\
  }\textbf {\bibinfo {volume} {47}},\ \bibinfo {pages} {415} (\bibinfo {year}
  {2000})}\BibitemShut {NoStop}%
\bibitem [{\citenamefont {Takamoto}\ \emph {et~al.}(2005)\citenamefont
  {Takamoto}, \citenamefont {Hong}, \citenamefont {Higashi},\ and\
  \citenamefont {Katori}}]{takamoto2005optical}%
  \BibitemOpen
  \bibfield  {author} {\bibinfo {author} {\bibfnamefont {M.}~\bibnamefont
  {Takamoto}}, \bibinfo {author} {\bibfnamefont {F.-L.}\ \bibnamefont {Hong}},
  \bibinfo {author} {\bibfnamefont {R.}~\bibnamefont {Higashi}},\ and\ \bibinfo
  {author} {\bibfnamefont {H.}~\bibnamefont {Katori}},\ }\bibfield  {title}
  {\bibinfo {title} {An optical lattice clock},\ }\href@noop {} {\bibfield
  {journal} {\bibinfo  {journal} {Nature}\ }\textbf {\bibinfo {volume} {435}},\
  \bibinfo {pages} {321} (\bibinfo {year} {2005})}\BibitemShut {NoStop}%
\bibitem [{\citenamefont {Ye}\ \emph {et~al.}(2008)\citenamefont {Ye},
  \citenamefont {Kimble},\ and\ \citenamefont {Katori}}]{ye2008quantum}%
  \BibitemOpen
  \bibfield  {author} {\bibinfo {author} {\bibfnamefont {J.}~\bibnamefont
  {Ye}}, \bibinfo {author} {\bibfnamefont {H.}~\bibnamefont {Kimble}},\ and\
  \bibinfo {author} {\bibfnamefont {H.}~\bibnamefont {Katori}},\ }\bibfield
  {title} {\bibinfo {title} {Quantum {S}tate {E}ngineering and {P}recision
  {M}etrology {U}sing {S}tate-{I}nsensitive {L}ight {T}raps},\ }\href@noop {}
  {\bibfield  {journal} {\bibinfo  {journal} {Science}\ }\textbf {\bibinfo
  {volume} {320}},\ \bibinfo {pages} {1734} (\bibinfo {year}
  {2008})}\BibitemShut {NoStop}%
\bibitem [{\citenamefont {Barber}\ \emph {et~al.}(2008)\citenamefont {Barber},
  \citenamefont {Stalnaker}, \citenamefont {Lemke}, \citenamefont {Poli},
  \citenamefont {Oates}, \citenamefont {Fortier}, \citenamefont {Diddams},
  \citenamefont {Hollberg}, \citenamefont {Hoyt}, \citenamefont {Taichenachev}
  \emph {et~al.}}]{barber2008optical}%
  \BibitemOpen
  \bibfield  {author} {\bibinfo {author} {\bibfnamefont {Z.~W.}\ \bibnamefont
  {Barber}}, \bibinfo {author} {\bibfnamefont {J.~E.}\ \bibnamefont
  {Stalnaker}}, \bibinfo {author} {\bibfnamefont {N.~D.}\ \bibnamefont
  {Lemke}}, \bibinfo {author} {\bibfnamefont {N.}~\bibnamefont {Poli}},
  \bibinfo {author} {\bibfnamefont {C.}~\bibnamefont {Oates}}, \bibinfo
  {author} {\bibfnamefont {T.}~\bibnamefont {Fortier}}, \bibinfo {author}
  {\bibfnamefont {S.}~\bibnamefont {Diddams}}, \bibinfo {author} {\bibfnamefont
  {L.}~\bibnamefont {Hollberg}}, \bibinfo {author} {\bibfnamefont
  {C.}~\bibnamefont {Hoyt}}, \bibinfo {author} {\bibfnamefont {A.}~\bibnamefont
  {Taichenachev}}, \emph {et~al.},\ }\bibfield  {title} {\bibinfo {title}
  {Optical {L}attice {I}nduced {L}ight {S}hifts in an {Y}b {A}tomic {C}lock},\
  }\href@noop {} {\bibfield  {journal} {\bibinfo  {journal} {Physical {R}eview
  {L}etters}\ }\textbf {\bibinfo {volume} {100}},\ \bibinfo {pages} {103002}
  (\bibinfo {year} {2008})}\BibitemShut {NoStop}%
\bibitem [{\citenamefont {Kohno}\ \emph {et~al.}(2009)\citenamefont {Kohno},
  \citenamefont {Yasuda}, \citenamefont {Hosaka}, \citenamefont {Inaba},
  \citenamefont {Nakajima},\ and\ \citenamefont {Hong}}]{kohno2009one}%
  \BibitemOpen
  \bibfield  {author} {\bibinfo {author} {\bibfnamefont {T.}~\bibnamefont
  {Kohno}}, \bibinfo {author} {\bibfnamefont {M.}~\bibnamefont {Yasuda}},
  \bibinfo {author} {\bibfnamefont {K.}~\bibnamefont {Hosaka}}, \bibinfo
  {author} {\bibfnamefont {H.}~\bibnamefont {Inaba}}, \bibinfo {author}
  {\bibfnamefont {Y.}~\bibnamefont {Nakajima}},\ and\ \bibinfo {author}
  {\bibfnamefont {F.-L.}\ \bibnamefont {Hong}},\ }\bibfield  {title} {\bibinfo
  {title} {One-{D}imensional {O}ptical {L}attice {C}lock with a {F}ermionic
  $^{171}${Y}b {I}sotope},\ }\href@noop {} {\bibfield  {journal} {\bibinfo
  {journal} {Applied Physics Express}\ }\textbf {\bibinfo {volume} {2}},\
  \bibinfo {pages} {072501} (\bibinfo {year} {2009})}\BibitemShut {NoStop}%
\bibitem [{\citenamefont {Westergaard}\ \emph {et~al.}(2011)\citenamefont
  {Westergaard}, \citenamefont {Lodewyck}, \citenamefont {Lorini},
  \citenamefont {Lecallier}, \citenamefont {Burt}, \citenamefont {Zawada},
  \citenamefont {Millo},\ and\ \citenamefont
  {Lemonde}}]{westergaard2011lattice}%
  \BibitemOpen
  \bibfield  {author} {\bibinfo {author} {\bibfnamefont {P.~G.}\ \bibnamefont
  {Westergaard}}, \bibinfo {author} {\bibfnamefont {J.}~\bibnamefont
  {Lodewyck}}, \bibinfo {author} {\bibfnamefont {L.}~\bibnamefont {Lorini}},
  \bibinfo {author} {\bibfnamefont {A.}~\bibnamefont {Lecallier}}, \bibinfo
  {author} {\bibfnamefont {E.}~\bibnamefont {Burt}}, \bibinfo {author}
  {\bibfnamefont {M.}~\bibnamefont {Zawada}}, \bibinfo {author} {\bibfnamefont
  {J.}~\bibnamefont {Millo}},\ and\ \bibinfo {author} {\bibfnamefont
  {P.}~\bibnamefont {Lemonde}},\ }\bibfield  {title} {\bibinfo {title}
  {Lattice-{I}nduced {F}requency {S}hifts in {S}r {O}ptical {L}attice {C}locks
  at the $10^{-17}$ {L}evel},\ }\href@noop {} {\bibfield  {journal} {\bibinfo
  {journal} {Physical {R}eview {L}etters}\ }\textbf {\bibinfo {volume} {106}},\
  \bibinfo {pages} {210801} (\bibinfo {year} {2011})}\BibitemShut {NoStop}%
\bibitem [{\citenamefont {Saskin}\ \emph {et~al.}(2019)\citenamefont {Saskin},
  \citenamefont {Wilson}, \citenamefont {Grinkemeyer},\ and\ \citenamefont
  {Thompson}}]{saskin2019narrow}%
  \BibitemOpen
  \bibfield  {author} {\bibinfo {author} {\bibfnamefont {S.}~\bibnamefont
  {Saskin}}, \bibinfo {author} {\bibfnamefont {J.}~\bibnamefont {Wilson}},
  \bibinfo {author} {\bibfnamefont {B.}~\bibnamefont {Grinkemeyer}},\ and\
  \bibinfo {author} {\bibfnamefont {J.}~\bibnamefont {Thompson}},\ }\bibfield
  {title} {\bibinfo {title} {Narrow-{L}ine {C}ooling and {I}maging of
  {Y}tterbium {A}toms in an {O}ptical {T}weezer {A}rray},\ }\href@noop {}
  {\bibfield  {journal} {\bibinfo  {journal} {Physical {R}eview {L}etters}\
  }\textbf {\bibinfo {volume} {122}},\ \bibinfo {pages} {143002} (\bibinfo
  {year} {2019})}\BibitemShut {NoStop}%
\bibitem [{\citenamefont {Ovsyannikov}\ \emph {et~al.}(2006)\citenamefont
  {Ovsyannikov}, \citenamefont {Pal'{c}hikov}, \citenamefont {Katori},\ and\
  \citenamefont {Takamoto}}]{ovsyannikov2006polarisation}%
  \BibitemOpen
  \bibfield  {author} {\bibinfo {author} {\bibfnamefont {V.~D.}\ \bibnamefont
  {Ovsyannikov}}, \bibinfo {author} {\bibfnamefont {V.~G.}\ \bibnamefont
  {Pal'{c}hikov}}, \bibinfo {author} {\bibfnamefont {H.}~\bibnamefont
  {Katori}},\ and\ \bibinfo {author} {\bibfnamefont {M.}~\bibnamefont
  {Takamoto}},\ }\bibfield  {title} {\bibinfo {title} {Polarisation and
  dispersion properties of light shifts in ultrastable optical frequency
  standards},\ }\href@noop {} {\bibfield  {journal} {\bibinfo  {journal}
  {Quantum Electronics}\ }\textbf {\bibinfo {volume} {36}} (\bibinfo {year}
  {2006})}\BibitemShut {NoStop}%
\bibitem [{\citenamefont {Mitroy}\ \emph {et~al.}(2010)\citenamefont {Mitroy},
  \citenamefont {Safronova},\ and\ \citenamefont {Clark}}]{mitroy2010theory}%
  \BibitemOpen
  \bibfield  {author} {\bibinfo {author} {\bibfnamefont {J.~A.}\ \bibnamefont
  {Mitroy}}, \bibinfo {author} {\bibfnamefont {M.~S.}\ \bibnamefont
  {Safronova}},\ and\ \bibinfo {author} {\bibfnamefont {C.~W.}\ \bibnamefont
  {Clark}},\ }\bibfield  {title} {\bibinfo {title} {Theory and applications of
  atomic and ionic polarizabilities},\ }\href@noop {} {\bibfield  {journal}
  {\bibinfo  {journal} {Journal of Physics B: Atomic, Molecular and Optical
  Physics}\ }\textbf {\bibinfo {volume} {43}},\ \bibinfo {pages} {202001}
  (\bibinfo {year} {2010})}\BibitemShut {NoStop}%
\bibitem [{\citenamefont {Brusch}\ \emph {et~al.}(2006)\citenamefont {Brusch},
  \citenamefont {Le~Targat}, \citenamefont {Baillard}, \citenamefont
  {Fouch{\'e}},\ and\ \citenamefont {Lemonde}}]{brusch2006hyperpolarizability}%
  \BibitemOpen
  \bibfield  {author} {\bibinfo {author} {\bibfnamefont {A.}~\bibnamefont
  {Brusch}}, \bibinfo {author} {\bibfnamefont {R.}~\bibnamefont {Le~Targat}},
  \bibinfo {author} {\bibfnamefont {X.}~\bibnamefont {Baillard}}, \bibinfo
  {author} {\bibfnamefont {M.}~\bibnamefont {Fouch{\'e}}},\ and\ \bibinfo
  {author} {\bibfnamefont {P.}~\bibnamefont {Lemonde}},\ }\bibfield  {title}
  {\bibinfo {title} {Hyperpolarizability {E}ffects in a {S}r {O}ptical
  {L}attice {C}lock},\ }\href@noop {} {\bibfield  {journal} {\bibinfo
  {journal} {Physical {R}eview {L}etters}\ }\textbf {\bibinfo {volume} {96}},\
  \bibinfo {pages} {103003} (\bibinfo {year} {2006})}\BibitemShut {NoStop}%
\bibitem [{\citenamefont {Katori}\ \emph {et~al.}(2009)\citenamefont {Katori},
  \citenamefont {Hashiguchi}, \citenamefont {Il’inova},\ and\ \citenamefont
  {Ovsiannikov}}]{katori2009magic}%
  \BibitemOpen
  \bibfield  {author} {\bibinfo {author} {\bibfnamefont {H.}~\bibnamefont
  {Katori}}, \bibinfo {author} {\bibfnamefont {K.}~\bibnamefont {Hashiguchi}},
  \bibinfo {author} {\bibfnamefont {E.~Y.}\ \bibnamefont {Il’inova}},\ and\
  \bibinfo {author} {\bibfnamefont {V.}~\bibnamefont {Ovsiannikov}},\
  }\bibfield  {title} {\bibinfo {title} {Magic {W}avelength to {M}ake {O}ptical
  {L}attice {C}locks {I}nsensitive to {A}tomic {M}otion},\ }\href@noop {}
  {\bibfield  {journal} {\bibinfo  {journal} {Physical {R}eview {L}etters}\
  }\textbf {\bibinfo {volume} {103}},\ \bibinfo {pages} {153004} (\bibinfo
  {year} {2009})}\BibitemShut {NoStop}%
\bibitem [{\citenamefont {Bloom}\ \emph {et~al.}(2014)\citenamefont {Bloom},
  \citenamefont {Nicholson}, \citenamefont {Williams}, \citenamefont
  {Campbell}, \citenamefont {Bishof}, \citenamefont {Zhang}, \citenamefont
  {Zhang}, \citenamefont {Bromley},\ and\ \citenamefont
  {Ye}}]{bloom2014optical}%
  \BibitemOpen
  \bibfield  {author} {\bibinfo {author} {\bibfnamefont {B.}~\bibnamefont
  {Bloom}}, \bibinfo {author} {\bibfnamefont {T.}~\bibnamefont {Nicholson}},
  \bibinfo {author} {\bibfnamefont {J.}~\bibnamefont {Williams}}, \bibinfo
  {author} {\bibfnamefont {S.}~\bibnamefont {Campbell}}, \bibinfo {author}
  {\bibfnamefont {M.}~\bibnamefont {Bishof}}, \bibinfo {author} {\bibfnamefont
  {X.}~\bibnamefont {Zhang}}, \bibinfo {author} {\bibfnamefont
  {W.}~\bibnamefont {Zhang}}, \bibinfo {author} {\bibfnamefont
  {S.}~\bibnamefont {Bromley}},\ and\ \bibinfo {author} {\bibfnamefont
  {J.}~\bibnamefont {Ye}},\ }\bibfield  {title} {\bibinfo {title} {An optical
  lattice clock with accuracy and stability at the $10^{-18}$ level},\
  }\href@noop {} {\bibfield  {journal} {\bibinfo  {journal} {Nature}\ }\textbf
  {\bibinfo {volume} {506}},\ \bibinfo {pages} {71} (\bibinfo {year}
  {2014})}\BibitemShut {NoStop}%
\bibitem [{\citenamefont {Brown}\ \emph {et~al.}(2017)\citenamefont {Brown},
  \citenamefont {Phillips}, \citenamefont {Beloy}, \citenamefont {McGrew},
  \citenamefont {Schioppo}, \citenamefont {Fasano}, \citenamefont {Milani},
  \citenamefont {Zhang}, \citenamefont {Hinkley}, \citenamefont {Leopardi}
  \emph {et~al.}}]{brown2017hyperpolarizability}%
  \BibitemOpen
  \bibfield  {author} {\bibinfo {author} {\bibfnamefont {R.~C.}\ \bibnamefont
  {Brown}}, \bibinfo {author} {\bibfnamefont {N.~B.}\ \bibnamefont {Phillips}},
  \bibinfo {author} {\bibfnamefont {K.}~\bibnamefont {Beloy}}, \bibinfo
  {author} {\bibfnamefont {W.~F.}\ \bibnamefont {McGrew}}, \bibinfo {author}
  {\bibfnamefont {M.}~\bibnamefont {Schioppo}}, \bibinfo {author}
  {\bibfnamefont {R.~J.}\ \bibnamefont {Fasano}}, \bibinfo {author}
  {\bibfnamefont {G.}~\bibnamefont {Milani}}, \bibinfo {author} {\bibfnamefont
  {X.}~\bibnamefont {Zhang}}, \bibinfo {author} {\bibfnamefont
  {N.}~\bibnamefont {Hinkley}}, \bibinfo {author} {\bibfnamefont
  {H.}~\bibnamefont {Leopardi}}, \emph {et~al.},\ }\bibfield  {title} {\bibinfo
  {title} {Hyperpolarizability and {O}perational {M}agic {W}avelength in an
  {O}ptical {L}attice {C}lock},\ }\href@noop {} {\bibfield  {journal} {\bibinfo
   {journal} {Physical {R}eview {L}etters}\ }\textbf {\bibinfo {volume}
  {119}},\ \bibinfo {pages} {253001} (\bibinfo {year} {2017})}\BibitemShut
  {NoStop}%
\bibitem [{\citenamefont {Porsev}\ \emph {et~al.}(2018)\citenamefont {Porsev},
  \citenamefont {Safronova}, \citenamefont {Safronova},\ and\ \citenamefont
  {Kozlov}}]{porsev2018multipolar}%
  \BibitemOpen
  \bibfield  {author} {\bibinfo {author} {\bibfnamefont {S.}~\bibnamefont
  {Porsev}}, \bibinfo {author} {\bibfnamefont {M.}~\bibnamefont {Safronova}},
  \bibinfo {author} {\bibfnamefont {U.}~\bibnamefont {Safronova}},\ and\
  \bibinfo {author} {\bibfnamefont {M.}~\bibnamefont {Kozlov}},\ }\bibfield
  {title} {\bibinfo {title} {Multipolar {P}olarizabilities and
  {H}yperpolarizabilities in the {S}r {O}ptical {L}attice {C}lock},\
  }\href@noop {} {\bibfield  {journal} {\bibinfo  {journal} {Physical {R}eview
  {L}etters}\ }\textbf {\bibinfo {volume} {120}},\ \bibinfo {pages} {063204}
  (\bibinfo {year} {2018})}\BibitemShut {NoStop}%
\bibitem [{\citenamefont {Ushijima}\ \emph {et~al.}(2018)\citenamefont
  {Ushijima}, \citenamefont {Takamoto},\ and\ \citenamefont
  {Katori}}]{ushijima2018operational}%
  \BibitemOpen
  \bibfield  {author} {\bibinfo {author} {\bibfnamefont {I.}~\bibnamefont
  {Ushijima}}, \bibinfo {author} {\bibfnamefont {M.}~\bibnamefont {Takamoto}},\
  and\ \bibinfo {author} {\bibfnamefont {H.}~\bibnamefont {Katori}},\
  }\bibfield  {title} {\bibinfo {title} {Operational {M}agic {I}ntensity for
  {S}r {O}ptical {L}attice {C}locks},\ }\href@noop {} {\bibfield  {journal}
  {\bibinfo  {journal} {Physical {R}eview {L}etters}\ }\textbf {\bibinfo
  {volume} {121}},\ \bibinfo {pages} {263202} (\bibinfo {year}
  {2018})}\BibitemShut {NoStop}%
\bibitem [{\citenamefont {Nemitz}\ \emph {et~al.}(2019)\citenamefont {Nemitz},
  \citenamefont {J{\o}rgensen}, \citenamefont {Yanagimoto}, \citenamefont
  {Bregolin},\ and\ \citenamefont {Katori}}]{nemitz2019modeling}%
  \BibitemOpen
  \bibfield  {author} {\bibinfo {author} {\bibfnamefont {N.}~\bibnamefont
  {Nemitz}}, \bibinfo {author} {\bibfnamefont {A.~A.}\ \bibnamefont
  {J{\o}rgensen}}, \bibinfo {author} {\bibfnamefont {R.}~\bibnamefont
  {Yanagimoto}}, \bibinfo {author} {\bibfnamefont {F.}~\bibnamefont
  {Bregolin}},\ and\ \bibinfo {author} {\bibfnamefont {H.}~\bibnamefont
  {Katori}},\ }\bibfield  {title} {\bibinfo {title} {Modeling light shifts in
  optical lattice clocks},\ }\href@noop {} {\bibfield  {journal} {\bibinfo
  {journal} {Physical {R}eview {A}}\ }\textbf {\bibinfo {volume} {99}},\
  \bibinfo {pages} {033424} (\bibinfo {year} {2019})}\BibitemShut {NoStop}%
\bibitem [{\citenamefont {Le~Targat}\ \emph {et~al.}(2012)\citenamefont
  {Le~Targat}, \citenamefont {Gartman}, \citenamefont {Lorini}, \citenamefont
  {Nag{\'o}rny}, \citenamefont {Gurov}, \citenamefont {Lemonde}, \citenamefont
  {Zawada},\ and\ \citenamefont {Lodewyck}}]{le2012comparison}%
  \BibitemOpen
  \bibfield  {author} {\bibinfo {author} {\bibfnamefont {R.}~\bibnamefont
  {Le~Targat}}, \bibinfo {author} {\bibfnamefont {R.}~\bibnamefont {Gartman}},
  \bibinfo {author} {\bibfnamefont {L.}~\bibnamefont {Lorini}}, \bibinfo
  {author} {\bibfnamefont {B.}~\bibnamefont {Nag{\'o}rny}}, \bibinfo {author}
  {\bibfnamefont {M.}~\bibnamefont {Gurov}}, \bibinfo {author} {\bibfnamefont
  {P.}~\bibnamefont {Lemonde}}, \bibinfo {author} {\bibfnamefont
  {M.}~\bibnamefont {Zawada}},\ and\ \bibinfo {author} {\bibfnamefont
  {J.}~\bibnamefont {Lodewyck}},\ }\bibfield  {title} {\bibinfo {title}
  {Comparison of two strontium optical lattice clocks in agreement at the
  $10^{-16}$ level},\ }in\ \href@noop {} {\emph {\bibinfo {booktitle} {2012
  IEEE International Frequency Control Symposium Proceedings}}}\ (\bibinfo
  {organization} {IEEE},\ \bibinfo {year} {2012})\ pp.\ \bibinfo {pages}
  {1--4}\BibitemShut {NoStop}%
\bibitem [{\citenamefont {Le~Targat}\ \emph {et~al.}(2013)\citenamefont
  {Le~Targat}, \citenamefont {Lorini}, \citenamefont {Le~Coq}, \citenamefont
  {Zawada}, \citenamefont {Gu{\'e}na}, \citenamefont {Abgrall}, \citenamefont
  {Gurov}, \citenamefont {Rosenbusch}, \citenamefont {Rovera}, \citenamefont
  {Nag{\'o}rny} \emph {et~al.}}]{le2013experimental}%
  \BibitemOpen
  \bibfield  {author} {\bibinfo {author} {\bibfnamefont {R.}~\bibnamefont
  {Le~Targat}}, \bibinfo {author} {\bibfnamefont {L.}~\bibnamefont {Lorini}},
  \bibinfo {author} {\bibfnamefont {Y.}~\bibnamefont {Le~Coq}}, \bibinfo
  {author} {\bibfnamefont {M.}~\bibnamefont {Zawada}}, \bibinfo {author}
  {\bibfnamefont {J.}~\bibnamefont {Gu{\'e}na}}, \bibinfo {author}
  {\bibfnamefont {M.}~\bibnamefont {Abgrall}}, \bibinfo {author} {\bibfnamefont
  {M.}~\bibnamefont {Gurov}}, \bibinfo {author} {\bibfnamefont
  {P.}~\bibnamefont {Rosenbusch}}, \bibinfo {author} {\bibfnamefont
  {D.}~\bibnamefont {Rovera}}, \bibinfo {author} {\bibfnamefont
  {B.}~\bibnamefont {Nag{\'o}rny}}, \emph {et~al.},\ }\bibfield  {title}
  {\bibinfo {title} {Experimental realization of an optical second with
  strontium lattice clocks},\ }\href@noop {} {\bibfield  {journal} {\bibinfo
  {journal} {Nature {C}ommunications}\ }\textbf {\bibinfo {volume} {4}},\
  \bibinfo {pages} {2109} (\bibinfo {year} {2013})}\BibitemShut {NoStop}%
\bibitem [{\citenamefont {Koller}\ \emph {et~al.}(2017)\citenamefont {Koller},
  \citenamefont {Grotti}, \citenamefont {Al-Masoudi}, \citenamefont
  {D{\"o}rscher}, \citenamefont {H{\"a}fner}, \citenamefont {Sterr},
  \citenamefont {Lisdat} \emph {et~al.}}]{koller2017transportable}%
  \BibitemOpen
  \bibfield  {author} {\bibinfo {author} {\bibfnamefont {S.}~\bibnamefont
  {Koller}}, \bibinfo {author} {\bibfnamefont {J.}~\bibnamefont {Grotti}},
  \bibinfo {author} {\bibfnamefont {A.}~\bibnamefont {Al-Masoudi}}, \bibinfo
  {author} {\bibfnamefont {S.}~\bibnamefont {D{\"o}rscher}}, \bibinfo {author}
  {\bibfnamefont {S.}~\bibnamefont {H{\"a}fner}}, \bibinfo {author}
  {\bibfnamefont {U.}~\bibnamefont {Sterr}}, \bibinfo {author} {\bibfnamefont
  {C.}~\bibnamefont {Lisdat}}, \emph {et~al.},\ }\bibfield  {title} {\bibinfo
  {title} {Transportable {O}ptical {L}attice {C}lock with 7$\times$ $10^{-17}$
  {U}ncertainty},\ }\href@noop {} {\bibfield  {journal} {\bibinfo  {journal}
  {Physical {R}eview {L}etters}\ }\textbf {\bibinfo {volume} {118}},\ \bibinfo
  {pages} {073601} (\bibinfo {year} {2017})}\BibitemShut {NoStop}%
\bibitem [{\citenamefont {Bloom}(2014)}]{bloom2014building}%
  \BibitemOpen
  \bibfield  {author} {\bibinfo {author} {\bibfnamefont {B.~J.}\ \bibnamefont
  {Bloom}},\ }\emph {\bibinfo {title} {Building a {B}etter {A}tomic {C}lock}},\
  \href@noop {} {Ph.D. thesis},\ \bibinfo  {school} {University of Colorado}
  (\bibinfo {year} {2014})\BibitemShut {NoStop}%
\bibitem [{\citenamefont {Zhou}\ \emph {et~al.}(2018)\citenamefont {Zhou},
  \citenamefont {Barthwal}, \citenamefont {Zhang}, \citenamefont {He},
  \citenamefont {Tang}, \citenamefont {Zhou}, \citenamefont {Wang},\ and\
  \citenamefont {Zhan}}]{zhou2018characterization}%
  \BibitemOpen
  \bibfield  {author} {\bibinfo {author} {\bibfnamefont {C.}~\bibnamefont
  {Zhou}}, \bibinfo {author} {\bibfnamefont {S.}~\bibnamefont {Barthwal}},
  \bibinfo {author} {\bibfnamefont {W.}~\bibnamefont {Zhang}}, \bibinfo
  {author} {\bibfnamefont {C.}~\bibnamefont {He}}, \bibinfo {author}
  {\bibfnamefont {B.}~\bibnamefont {Tang}}, \bibinfo {author} {\bibfnamefont
  {L.}~\bibnamefont {Zhou}}, \bibinfo {author} {\bibfnamefont {J.}~\bibnamefont
  {Wang}},\ and\ \bibinfo {author} {\bibfnamefont {M.-S.}\ \bibnamefont
  {Zhan}},\ }\bibfield  {title} {\bibinfo {title} {Characterization and
  optimization of a tapered amplifier by its spectra through a long multi-pass
  rubidium absorption cell},\ }\href@noop {} {\bibfield  {journal} {\bibinfo
  {journal} {Applied optics}\ }\textbf {\bibinfo {volume} {57}},\ \bibinfo
  {pages} {7427} (\bibinfo {year} {2018})}\BibitemShut {NoStop}%
\bibitem [{\citenamefont {Voigt}\ \emph {et~al.}(2001)\citenamefont {Voigt},
  \citenamefont {Schilder}, \citenamefont {Spreeuw},\ and\ \citenamefont {Van
  Den~Heuvell}}]{voigt2001characterization}%
  \BibitemOpen
  \bibfield  {author} {\bibinfo {author} {\bibfnamefont {D.}~\bibnamefont
  {Voigt}}, \bibinfo {author} {\bibfnamefont {E.}~\bibnamefont {Schilder}},
  \bibinfo {author} {\bibfnamefont {R.}~\bibnamefont {Spreeuw}},\ and\ \bibinfo
  {author} {\bibfnamefont {H.~V.~L.}\ \bibnamefont {Van Den~Heuvell}},\
  }\bibfield  {title} {\bibinfo {title} {Characterization of a high-power
  tapered semiconductor amplifier system},\ }\href@noop {} {\bibfield
  {journal} {\bibinfo  {journal} {Applied Physics B}\ }\textbf {\bibinfo
  {volume} {72}},\ \bibinfo {pages} {279} (\bibinfo {year} {2001})}\BibitemShut
  {NoStop}%
\bibitem [{\citenamefont {Bolpasi}\ and\ \citenamefont
  {Von~Klitzing}(2010)}]{bolpasi2010double}%
  \BibitemOpen
  \bibfield  {author} {\bibinfo {author} {\bibfnamefont {V.}~\bibnamefont
  {Bolpasi}}\ and\ \bibinfo {author} {\bibfnamefont {W.}~\bibnamefont
  {Von~Klitzing}},\ }\bibfield  {title} {\bibinfo {title} {Double-pass tapered
  amplifier diode laser with an output power of 1 {W} for an injection power of
  only 200 $\mu${W}},\ }\href@noop {} {\bibfield  {journal} {\bibinfo
  {journal} {Review of Scientific Instruments}\ }\textbf {\bibinfo {volume}
  {81}},\ \bibinfo {pages} {113108} (\bibinfo {year} {2010})}\BibitemShut
  {NoStop}%
\bibitem [{\citenamefont {Nyman}\ \emph {et~al.}(2006)\citenamefont {Nyman},
  \citenamefont {Varoquaux}, \citenamefont {Villier}, \citenamefont {Sacchet},
  \citenamefont {Moron}, \citenamefont {Le~Coq}, \citenamefont {Aspect},\ and\
  \citenamefont {Bouyer}}]{nyman2006tapered}%
  \BibitemOpen
  \bibfield  {author} {\bibinfo {author} {\bibfnamefont {R.~A.}\ \bibnamefont
  {Nyman}}, \bibinfo {author} {\bibfnamefont {G.}~\bibnamefont {Varoquaux}},
  \bibinfo {author} {\bibfnamefont {B.}~\bibnamefont {Villier}}, \bibinfo
  {author} {\bibfnamefont {D.}~\bibnamefont {Sacchet}}, \bibinfo {author}
  {\bibfnamefont {F.}~\bibnamefont {Moron}}, \bibinfo {author} {\bibfnamefont
  {Y.}~\bibnamefont {Le~Coq}}, \bibinfo {author} {\bibfnamefont
  {A.}~\bibnamefont {Aspect}},\ and\ \bibinfo {author} {\bibfnamefont
  {P.}~\bibnamefont {Bouyer}},\ }\bibfield  {title} {\bibinfo {title}
  {Tapered-amplified antireflection-coated laser diodes for potassium and
  rubidium atomic-physics experiments},\ }\href@noop {} {\bibfield  {journal}
  {\bibinfo  {journal} {Review of Scientific Instruments}\ }\textbf {\bibinfo
  {volume} {77}},\ \bibinfo {pages} {033105} (\bibinfo {year}
  {2006})}\BibitemShut {NoStop}%
\bibitem [{\citenamefont {Bienaim{\'e}}\ \emph {et~al.}(2012)\citenamefont
  {Bienaim{\'e}}, \citenamefont {Barontini}, \citenamefont {de~L{\'e}pinay},
  \citenamefont {Bellando}, \citenamefont {Chab{\'e}},\ and\ \citenamefont
  {Kaiser}}]{bienaime2012fast}%
  \BibitemOpen
  \bibfield  {author} {\bibinfo {author} {\bibfnamefont {T.}~\bibnamefont
  {Bienaim{\'e}}}, \bibinfo {author} {\bibfnamefont {G.}~\bibnamefont
  {Barontini}}, \bibinfo {author} {\bibfnamefont {L.~M.}\ \bibnamefont
  {de~L{\'e}pinay}}, \bibinfo {author} {\bibfnamefont {L.}~\bibnamefont
  {Bellando}}, \bibinfo {author} {\bibfnamefont {J.}~\bibnamefont
  {Chab{\'e}}},\ and\ \bibinfo {author} {\bibfnamefont {R.}~\bibnamefont
  {Kaiser}},\ }\bibfield  {title} {\bibinfo {title} {Fast compression of a cold
  atomic cloud using a blue-detuned crossed dipole trap},\ }\href@noop {}
  {\bibfield  {journal} {\bibinfo  {journal} {Physical Review A}\ }\textbf
  {\bibinfo {volume} {86}},\ \bibinfo {pages} {053412} (\bibinfo {year}
  {2012})}\BibitemShut {NoStop}%
\bibitem [{\citenamefont {Dumke}\ \emph {et~al.}(2002)\citenamefont {Dumke},
  \citenamefont {Volk}, \citenamefont {M{\"u}ther}, \citenamefont {Buchkremer},
  \citenamefont {Birkl},\ and\ \citenamefont {Ertmer}}]{dumke2002micro}%
  \BibitemOpen
  \bibfield  {author} {\bibinfo {author} {\bibfnamefont {R.}~\bibnamefont
  {Dumke}}, \bibinfo {author} {\bibfnamefont {M.}~\bibnamefont {Volk}},
  \bibinfo {author} {\bibfnamefont {T.}~\bibnamefont {M{\"u}ther}}, \bibinfo
  {author} {\bibfnamefont {F.}~\bibnamefont {Buchkremer}}, \bibinfo {author}
  {\bibfnamefont {G.}~\bibnamefont {Birkl}},\ and\ \bibinfo {author}
  {\bibfnamefont {W.}~\bibnamefont {Ertmer}},\ }\bibfield  {title} {\bibinfo
  {title} {Micro-optical realization of arrays of selectively addressable
  dipole traps: a scalable configuration for quantum computation with atomic
  qubits},\ }\href@noop {} {\bibfield  {journal} {\bibinfo  {journal} {Physical
  Review Letters}\ }\textbf {\bibinfo {volume} {89}},\ \bibinfo {pages}
  {097903} (\bibinfo {year} {2002})}\BibitemShut {NoStop}%
\bibitem [{\citenamefont {Takamoto}\ \emph {et~al.}(2020)\citenamefont
  {Takamoto}, \citenamefont {Ushijima}, \citenamefont {Ohmae}, \citenamefont
  {Yahagi}, \citenamefont {Kokado}, \citenamefont {Shinkai},\ and\
  \citenamefont {Katori}}]{takamoto2020test}%
  \BibitemOpen
  \bibfield  {author} {\bibinfo {author} {\bibfnamefont {M.}~\bibnamefont
  {Takamoto}}, \bibinfo {author} {\bibfnamefont {I.}~\bibnamefont {Ushijima}},
  \bibinfo {author} {\bibfnamefont {N.}~\bibnamefont {Ohmae}}, \bibinfo
  {author} {\bibfnamefont {T.}~\bibnamefont {Yahagi}}, \bibinfo {author}
  {\bibfnamefont {K.}~\bibnamefont {Kokado}}, \bibinfo {author} {\bibfnamefont
  {H.}~\bibnamefont {Shinkai}},\ and\ \bibinfo {author} {\bibfnamefont
  {H.}~\bibnamefont {Katori}},\ }\bibfield  {title} {\bibinfo {title} {Test of
  general relativity by a pair of transportable optical lattice clocks},\
  }\href@noop {} {\bibfield  {journal} {\bibinfo  {journal} {Nature Photonics}\
  ,\ \bibinfo {pages} {1}} (\bibinfo {year} {2020})}\BibitemShut {NoStop}%
\bibitem [{\citenamefont {Origlia}\ \emph {et~al.}(2018)\citenamefont
  {Origlia}, \citenamefont {Pramod}, \citenamefont {Schiller}, \citenamefont
  {Singh}, \citenamefont {Bongs}, \citenamefont {Schwarz}, \citenamefont
  {Al-Masoudi}, \citenamefont {D{\"o}rscher}, \citenamefont {Herbers},
  \citenamefont {H{\"a}fner} \emph {et~al.}}]{origlia2018towards}%
  \BibitemOpen
  \bibfield  {author} {\bibinfo {author} {\bibfnamefont {S.}~\bibnamefont
  {Origlia}}, \bibinfo {author} {\bibfnamefont {M.~S.}\ \bibnamefont {Pramod}},
  \bibinfo {author} {\bibfnamefont {S.}~\bibnamefont {Schiller}}, \bibinfo
  {author} {\bibfnamefont {Y.}~\bibnamefont {Singh}}, \bibinfo {author}
  {\bibfnamefont {K.}~\bibnamefont {Bongs}}, \bibinfo {author} {\bibfnamefont
  {R.}~\bibnamefont {Schwarz}}, \bibinfo {author} {\bibfnamefont
  {A.}~\bibnamefont {Al-Masoudi}}, \bibinfo {author} {\bibfnamefont
  {S.}~\bibnamefont {D{\"o}rscher}}, \bibinfo {author} {\bibfnamefont
  {S.}~\bibnamefont {Herbers}}, \bibinfo {author} {\bibfnamefont
  {S.}~\bibnamefont {H{\"a}fner}}, \emph {et~al.},\ }\bibfield  {title}
  {\bibinfo {title} {Towards an optical clock for space: Compact,
  high-performance optical lattice clock based on bosonic atoms},\ }\href@noop
  {} {\bibfield  {journal} {\bibinfo  {journal} {Physical Review A}\ }\textbf
  {\bibinfo {volume} {98}},\ \bibinfo {pages} {053443} (\bibinfo {year}
  {2018})}\BibitemShut {NoStop}%
\bibitem [{Note1()}]{Note1}%
  \BibitemOpen
  \bibinfo {note} {Equipment used for these measurements includes: Ti:S laser
  (M-Squared SolsTiS), amplified diode laser (Toptica TA Pro), and optical
  spectrum analyzer (Yokogawa AQ6373B). Product information is provided for
  informational purposes and does not represent an endorsement by
  NIST.}\BibitemShut {Stop}%
\bibitem [{\citenamefont {McGrew}\ \emph {et~al.}(2018)\citenamefont {McGrew},
  \citenamefont {Zhang}, \citenamefont {Fasano}, \citenamefont {Sch{\"a}ffer},
  \citenamefont {Beloy}, \citenamefont {Nicolodi}, \citenamefont {Brown},
  \citenamefont {Hinkley}, \citenamefont {Milani}, \citenamefont {Schioppo}
  \emph {et~al.}}]{mcgrew2018atomic}%
  \BibitemOpen
  \bibfield  {author} {\bibinfo {author} {\bibfnamefont {W.}~\bibnamefont
  {McGrew}}, \bibinfo {author} {\bibfnamefont {X.}~\bibnamefont {Zhang}},
  \bibinfo {author} {\bibfnamefont {R.}~\bibnamefont {Fasano}}, \bibinfo
  {author} {\bibfnamefont {S.}~\bibnamefont {Sch{\"a}ffer}}, \bibinfo {author}
  {\bibfnamefont {K.}~\bibnamefont {Beloy}}, \bibinfo {author} {\bibfnamefont
  {D.}~\bibnamefont {Nicolodi}}, \bibinfo {author} {\bibfnamefont
  {R.}~\bibnamefont {Brown}}, \bibinfo {author} {\bibfnamefont
  {N.}~\bibnamefont {Hinkley}}, \bibinfo {author} {\bibfnamefont
  {G.}~\bibnamefont {Milani}}, \bibinfo {author} {\bibfnamefont
  {M.}~\bibnamefont {Schioppo}}, \emph {et~al.},\ }\bibfield  {title} {\bibinfo
  {title} {Atomic clock performance enabling geodesy below the centimetre
  level},\ }\href@noop {} {\bibfield  {journal} {\bibinfo  {journal} {Nature}\
  }\textbf {\bibinfo {volume} {564}},\ \bibinfo {pages} {87} (\bibinfo {year}
  {2018})}\BibitemShut {NoStop}%
\bibitem [{\citenamefont {Bakker}\ \emph {et~al.}(1999)\citenamefont {Bakker},
  \citenamefont {Freriks},\ and\ \citenamefont {Kroesen}}]{bakker1999new}%
  \BibitemOpen
  \bibfield  {author} {\bibinfo {author} {\bibfnamefont {L.}~\bibnamefont
  {Bakker}}, \bibinfo {author} {\bibfnamefont {J.}~\bibnamefont {Freriks}},\
  and\ \bibinfo {author} {\bibfnamefont {G.}~\bibnamefont {Kroesen}},\
  }\bibfield  {title} {\bibinfo {title} {A new {ASE} filter: the 20-fold prism
  monochromator},\ }\href@noop {} {\bibfield  {journal} {\bibinfo  {journal}
  {Measurement Science and Technology}\ }\textbf {\bibinfo {volume} {10}},\
  \bibinfo {pages} {L25} (\bibinfo {year} {1999})}\BibitemShut {NoStop}%
\bibitem [{\citenamefont {Marling}\ \emph {et~al.}(1979)\citenamefont
  {Marling}, \citenamefont {Nilsen}, \citenamefont {West},\ and\ \citenamefont
  {Wood}}]{marling1979ultrahigh}%
  \BibitemOpen
  \bibfield  {author} {\bibinfo {author} {\bibfnamefont {J.~B.}\ \bibnamefont
  {Marling}}, \bibinfo {author} {\bibfnamefont {J.}~\bibnamefont {Nilsen}},
  \bibinfo {author} {\bibfnamefont {L.}~\bibnamefont {West}},\ and\ \bibinfo
  {author} {\bibfnamefont {L.}~\bibnamefont {Wood}},\ }\bibfield  {title}
  {\bibinfo {title} {An ultrahigh-{Q} isotropically sensitive optical filter
  employing atomic resonance transitions},\ }\href@noop {} {\bibfield
  {journal} {\bibinfo  {journal} {Journal of Applied Physics}\ }\textbf
  {\bibinfo {volume} {50}},\ \bibinfo {pages} {610} (\bibinfo {year}
  {1979})}\BibitemShut {NoStop}%
\bibitem [{\citenamefont {Magnusson}\ and\ \citenamefont
  {Wang}(1992)}]{magnusson1992new}%
  \BibitemOpen
  \bibfield  {author} {\bibinfo {author} {\bibfnamefont {R.}~\bibnamefont
  {Magnusson}}\ and\ \bibinfo {author} {\bibfnamefont {S.}~\bibnamefont
  {Wang}},\ }\bibfield  {title} {\bibinfo {title} {New principle for optical
  filters},\ }\href@noop {} {\bibfield  {journal} {\bibinfo  {journal} {Applied
  {P}hysics {L}etters}\ }\textbf {\bibinfo {volume} {61}},\ \bibinfo {pages}
  {1022} (\bibinfo {year} {1992})}\BibitemShut {NoStop}%
\bibitem [{\citenamefont {Glebov}\ \emph {et~al.}(2012)\citenamefont {Glebov},
  \citenamefont {Mokhun}, \citenamefont {Rapaport}, \citenamefont {Vergnole},
  \citenamefont {Smirnov},\ and\ \citenamefont {Glebov}}]{glebov2012volume}%
  \BibitemOpen
  \bibfield  {author} {\bibinfo {author} {\bibfnamefont {A.~L.}\ \bibnamefont
  {Glebov}}, \bibinfo {author} {\bibfnamefont {O.}~\bibnamefont {Mokhun}},
  \bibinfo {author} {\bibfnamefont {A.}~\bibnamefont {Rapaport}}, \bibinfo
  {author} {\bibfnamefont {S.}~\bibnamefont {Vergnole}}, \bibinfo {author}
  {\bibfnamefont {V.}~\bibnamefont {Smirnov}},\ and\ \bibinfo {author}
  {\bibfnamefont {L.~B.}\ \bibnamefont {Glebov}},\ }\bibfield  {title}
  {\bibinfo {title} {Volume {B}ragg gratings as ultra-narrow and multiband
  optical filters},\ }in\ \href@noop {} {\emph {\bibinfo {booktitle}
  {Micro-Optics 2012}}},\ Vol.\ \bibinfo {volume} {8428}\ (\bibinfo
  {organization} {International Society for Optics and Photonics},\ \bibinfo
  {year} {2012})\ p.\ \bibinfo {pages} {84280C}\BibitemShut {NoStop}%
\bibitem [{\citenamefont {Lumeau}\ \emph {et~al.}(2006)\citenamefont {Lumeau},
  \citenamefont {Glebov},\ and\ \citenamefont {Smirnov}}]{lumeau2006tunable}%
  \BibitemOpen
  \bibfield  {author} {\bibinfo {author} {\bibfnamefont {J.}~\bibnamefont
  {Lumeau}}, \bibinfo {author} {\bibfnamefont {L.}~\bibnamefont {Glebov}},\
  and\ \bibinfo {author} {\bibfnamefont {V.}~\bibnamefont {Smirnov}},\
  }\bibfield  {title} {\bibinfo {title} {Tunable narrowband filter based on a
  combination of {F}abry-{P}erot etalon and volume {B}ragg grating},\
  }\href@noop {} {\bibfield  {journal} {\bibinfo  {journal} {Optics {L}etters}\
  }\textbf {\bibinfo {volume} {31}},\ \bibinfo {pages} {2417} (\bibinfo {year}
  {2006})}\BibitemShut {NoStop}%
\bibitem [{\citenamefont {Hellstrom}\ \emph {et~al.}(2007)\citenamefont
  {Hellstrom}, \citenamefont {Jacobsson}, \citenamefont {Pasiskevicius},\ and\
  \citenamefont {Laurell}}]{hellstrom2007finite}%
  \BibitemOpen
  \bibfield  {author} {\bibinfo {author} {\bibfnamefont {J.~E.}\ \bibnamefont
  {Hellstrom}}, \bibinfo {author} {\bibfnamefont {B.}~\bibnamefont
  {Jacobsson}}, \bibinfo {author} {\bibfnamefont {V.}~\bibnamefont
  {Pasiskevicius}},\ and\ \bibinfo {author} {\bibfnamefont {F.}~\bibnamefont
  {Laurell}},\ }\bibfield  {title} {\bibinfo {title} {Finite beams in
  reflective volume {B}ragg gratings: theory and experiments},\ }\href@noop {}
  {\bibfield  {journal} {\bibinfo  {journal} {IEEE Journal of {Q}uantum
  {E}lectronics}\ }\textbf {\bibinfo {volume} {44}},\ \bibinfo {pages} {81}
  (\bibinfo {year} {2007})}\BibitemShut {NoStop}%
\bibitem [{Note2()}]{Note2}%
  \BibitemOpen
  \bibinfo {note} {Because the heterodyne measurements include ASE spectrum
  from both positive and negative heterodyne frequencies, essentially doubling
  the expected measurement of the ASE amplitude, comparison of the observed
  spectrum in this frequency region should be made relative to the interpolated
  Gaussian fit plus an additional 6 dB.}\BibitemShut {Stop}%
\bibitem [{\citenamefont {Ott}\ \emph {et~al.}(2015)\citenamefont {Ott},
  \citenamefont {SeGall}, \citenamefont {Divliansky}, \citenamefont {Venus},\
  and\ \citenamefont {Glebov}}]{ott2015high}%
  \BibitemOpen
  \bibfield  {author} {\bibinfo {author} {\bibfnamefont {D.}~\bibnamefont
  {Ott}}, \bibinfo {author} {\bibfnamefont {M.}~\bibnamefont {SeGall}},
  \bibinfo {author} {\bibfnamefont {I.}~\bibnamefont {Divliansky}}, \bibinfo
  {author} {\bibfnamefont {G.}~\bibnamefont {Venus}},\ and\ \bibinfo {author}
  {\bibfnamefont {L.}~\bibnamefont {Glebov}},\ }\bibfield  {title} {\bibinfo
  {title} {High-contrast filtering by multipass diffraction between paired
  volume {B}ragg gratings},\ }\href@noop {} {\bibfield  {journal} {\bibinfo
  {journal} {Applied {O}ptics}\ }\textbf {\bibinfo {volume} {54}},\ \bibinfo
  {pages} {9065} (\bibinfo {year} {2015})}\BibitemShut {NoStop}%
\bibitem [{\citenamefont {Ciapurin}\ \emph {et~al.}(2005)\citenamefont
  {Ciapurin}, \citenamefont {Glebov},\ and\ \citenamefont
  {Smirnov}}]{ciapurin2005modeling}%
  \BibitemOpen
  \bibfield  {author} {\bibinfo {author} {\bibfnamefont {I.~V.}\ \bibnamefont
  {Ciapurin}}, \bibinfo {author} {\bibfnamefont {L.~B.}\ \bibnamefont
  {Glebov}},\ and\ \bibinfo {author} {\bibfnamefont {V.~I.}\ \bibnamefont
  {Smirnov}},\ }\bibfield  {title} {\bibinfo {title} {Modeling of {G}aussian
  beam diffraction on volume {B}ragg gratings in {PTR} glass},\ }in\ \href@noop
  {} {\emph {\bibinfo {booktitle} {Practical Holography XIX: Materials and
  Applications}}},\ Vol.\ \bibinfo {volume} {5742}\ (\bibinfo {organization}
  {International Society for Optics and Photonics},\ \bibinfo {year} {2005})\
  pp.\ \bibinfo {pages} {183--194}\BibitemShut {NoStop}%
\bibitem [{\citenamefont {Kondov}\ \emph {et~al.}(2019)\citenamefont {Kondov},
  \citenamefont {Lee}, \citenamefont {Leung}, \citenamefont {Liedl},
  \citenamefont {Majewska}, \citenamefont {Moszynski},\ and\ \citenamefont
  {Zelevinsky}}]{kondov2019molecular}%
  \BibitemOpen
  \bibfield  {author} {\bibinfo {author} {\bibfnamefont {S.}~\bibnamefont
  {Kondov}}, \bibinfo {author} {\bibfnamefont {C.-H.}\ \bibnamefont {Lee}},
  \bibinfo {author} {\bibfnamefont {K.}~\bibnamefont {Leung}}, \bibinfo
  {author} {\bibfnamefont {C.}~\bibnamefont {Liedl}}, \bibinfo {author}
  {\bibfnamefont {I.}~\bibnamefont {Majewska}}, \bibinfo {author}
  {\bibfnamefont {R.}~\bibnamefont {Moszynski}},\ and\ \bibinfo {author}
  {\bibfnamefont {T.}~\bibnamefont {Zelevinsky}},\ }\bibfield  {title}
  {\bibinfo {title} {Molecular lattice clock with long vibrational coherence},\
  }\href@noop {} {\bibfield  {journal} {\bibinfo  {journal} {Nature Physics}\
  }\textbf {\bibinfo {volume} {15}},\ \bibinfo {pages} {1118} (\bibinfo {year}
  {2019})}\BibitemShut {NoStop}%
\bibitem [{\citenamefont {Zhan}\ \emph {et~al.}(2015)\citenamefont {Zhan},
  \citenamefont {Duan}, \citenamefont {Zhou}, \citenamefont {Yao},
  \citenamefont {Xu},\ and\ \citenamefont {Hu}}]{zhan2015investigating}%
  \BibitemOpen
  \bibfield  {author} {\bibinfo {author} {\bibfnamefont {S.}~\bibnamefont
  {Zhan}}, \bibinfo {author} {\bibfnamefont {X.-C.}\ \bibnamefont {Duan}},
  \bibinfo {author} {\bibfnamefont {M.-K.}\ \bibnamefont {Zhou}}, \bibinfo
  {author} {\bibfnamefont {H.-B.}\ \bibnamefont {Yao}}, \bibinfo {author}
  {\bibfnamefont {W.-J.}\ \bibnamefont {Xu}},\ and\ \bibinfo {author}
  {\bibfnamefont {Z.-K.}\ \bibnamefont {Hu}},\ }\bibfield  {title} {\bibinfo
  {title} {Investigating the frequency-dependent amplification of a tapered
  amplifier in atom interferometers},\ }\href@noop {} {\bibfield  {journal}
  {\bibinfo  {journal} {Optics Letters}\ }\textbf {\bibinfo {volume} {40}},\
  \bibinfo {pages} {29} (\bibinfo {year} {2015})}\BibitemShut {NoStop}%
\bibitem [{\citenamefont {Shortt}\ \emph {et~al.}(2019)\citenamefont {Shortt},
  \citenamefont {Mondin}, \citenamefont {McNamara}, \citenamefont {Dahl},
  \citenamefont {Lecomte},\ and\ \citenamefont {Duque}}]{shortt2019lisa}%
  \BibitemOpen
  \bibfield  {author} {\bibinfo {author} {\bibfnamefont {B.}~\bibnamefont
  {Shortt}}, \bibinfo {author} {\bibfnamefont {L.}~\bibnamefont {Mondin}},
  \bibinfo {author} {\bibfnamefont {P.}~\bibnamefont {McNamara}}, \bibinfo
  {author} {\bibfnamefont {K.}~\bibnamefont {Dahl}}, \bibinfo {author}
  {\bibfnamefont {S.}~\bibnamefont {Lecomte}},\ and\ \bibinfo {author}
  {\bibfnamefont {P.}~\bibnamefont {Duque}},\ }\bibfield  {title} {\bibinfo
  {title} {LISA laser system and European development strategy},\ }in\
  \href@noop {} {\emph {\bibinfo {booktitle} {International Conference on Space
  Optics 2018}}},\ Vol.\ \bibinfo {volume} {11180}\ (\bibinfo
  {organization} {International Society for Optics and Photonics},\ \bibinfo
  {year} {2019})\ p.\ \bibinfo {pages} {111800D}\BibitemShut {NoStop}%
\end{thebibliography}

\providecommand{\noopsort}[1]{}\providecommand{\singleletter}[1]{#1}%
%

\end{document}